\documentclass[a4paper,12pt]{article}
\usepackage{epsfig}
\usepackage{latexsym}
\usepackage{graphicx}
\usepackage{amsfonts}
\usepackage{amsmath}
\usepackage{natbib}
\usepackage{xcolor}
\usepackage[hyperindex=false,citecolor=black,linkcolor=black,linkbordercolor={1 1 1},colorlinks=false,citebordercolor={1 1 1}]{hyperref}
\usepackage[blocks]{authblk}
\usepackage{longtable}

\date{}

\title{\bf DMAW 2010 LEGACY \\  {\it the Presentation Review}:  \\ Dark Matter in Galaxies \\  {\it with its}   Explanatory  Notes }

\author[1]{\bf Paolo Salucci} 
\author[2]{\bf Christiane Frigerio Martins} 
\author[3]{\bf Andrea Lapi} 
\affil[1]{\small SISSA/ISAS, via Bonomea 265, 34136 Trieste, Italy}
\affil[2]{\small INFES/Univ. Federal Fluminense, av. Jo\~{a}o Jazbik, 28470 Santo Ant\^{o}nio de P\'{a}dua, RJ, Brazil}
\affil[3] {\small Dip. Fisica, Univ. di Roma "Tor Vergata", Via Ricerca Scientifica 1, 00133 Roma, Italy }

\begin{document} 
\maketitle

 {\bf With the scientific collaboration of Elena Aprile, 
\author{Mariangela Bernardi, Albert  Bosma, Erwin de Blok,  
Ken Freeman, Refael Gavazzi, Gianfranco Gentile, Gerry  Gilmore,
Uli Klein, Gary Mamon, Claudia Maraston, Nicola Napolitano, Pierre Salati, 
Chiara Tonini,  Mark  Wilkinson, Irina Yegorova.}}

\

With the  support and encouragement  of \author{A. Burkert,  J. Bailin,
  P. Biermann, A. Bressan, S. Capozziello,  L. Danese, C. Frenk,  S. Leach , M. Roos, V. Rubin}

\section{General Introduction}

{\small The Journal Club Talk "Dark Matter in Galaxies" was delivered within  the "Dark Matter Awareness Week (1-8 December 2010)" at 140 institutes in 46 countries and was followed by 4200 people. More information on DMAW can be found at its website: www.sissa.it/ap/dmg/index.html and a short movie that documents this worldwide initiative at: \url{http://www.youtube.com/watch?v=AOBit8a-1Fw}.

For this DMAW 2010 talk  there was a standard reference presentation with explaining notes  prepared by a coordinated  pool of leading scientists in the field,
which the various speakers  adapted and delivered in their Institutes so as to suit local circumstances, audiences and their own personal preferences.
Importantly, the speakers  subsequently provided us with comments on the reference material  and  on  the initiative itself (see Section 1.3 to download them). 

After the event, in response to participants' feedback, we improved and upgraded the guideline material to a "Presentation Review" on the subject "Dark Matter in Galaxies", for which we provide a .pptx file and the related Explanatory Notes.
This is described below together with suggestions of who might be interested in using it and how to do it.}

\subsection {What is the Presentation Review + Explanatory Notes?}

{\small This is an innovative scientific product, introduced for fostering scientific debate on Cosmology and  for rapidly diffusing, among the scientific community, results in a field where,  we believe,  knowledge is rapidly growing.   Supported by the   large number of speakers delivering the DMAW 2010 talk,  by  their reactions and by  those of  their numerous audiences,  we decided  to address    the whole  community of cosmologists and astrophysicists. We  prepared  this Presentation Review, convinced that there is  a generalized request of knowledge about the DM mystery at galactic scale.

The Presentation is a Review Talk at post-graduate level, (2 x 45) minutes long, on the specific subject: "The distribution of Dark (and Luminous) Matter in Galaxies", although other cosmological issues (e.g. galaxy formation) are treated, but from the phenomenological perspective of the galaxy mass distribution.

The Presentation Review is not a PowerPoint translation of an already existing Review Paper, but it has been prepared directly as a Presentation on the above described subject. The Explanatory Notes  are not the written version of the Presentation, but concepts, formulas and references that help with
understanding the slides and preparing an eventual speaker to comment  on them to an audience.

 In detail, their scope is to enable  the user who reads them to  follow the Presentation easily and to get the greatest benefit from it. The user who,  instead,  chooses  to  carefully {\it study}  them, would be able herself/himself to deliver the Review  Talk. For this reason, there is some additional material in the Notes (arguments, references, formulas) for the benefit of a prospective speaker or any user who wants to deepen the  knowledge on the subject. 

The presentation runs also  two movies in prefixed slides. These movies  are especially devised for the Presentation and concern the Cosmological Simulations  and The Universal Rotation Curve of galaxies.

It may be obvious, however,  it could still  be  worth to stress that the  Explanatory Notes without  the .ppt Presentation are worthless. On the other hand,  the Presentation Review file without the Notes is difficult to ``use''. 

An other  clarification maybe necessary: we are envisaging this Review as being one section of a virtual course on ``Physical Cosmology'' organized as follows: Big Bang Cosmology (3 x 45'), Structure formation (4 x 45'), Dark Ages and objects at high redshifts (3 x 45'), The distribution of DM in Groups, Clusters and larger scales (2 x 45'), The distribution of DM in Galaxies (2 x 45'), The physics of baryons in forming the luminous structure of  Universe (4 x 45'), Candidates for the DM particle and their detection (3 x 45'), Alternatives to Dark Matter (1 x 45'). We have not  written out this programme  to anticipate a future project, but to frame this Review Talk in its  Cosmological Context. More specifically, our Review aims to bring the user up to a good level of knowledge of its core subject: the Dark and Luminous matter distribution in galaxies. For other topics, which are occasionally touched on by the review, the user is redirected to more specialized reviews, lecture notes and books.  And, of course, a number of issues involving DM  will not be dealt in this Review.

Let us also stress that, contrary to the situation for written reviews, with this one it is up to the users, and not only the authors, to establish the relative and absolute importance of each statement, formula, reference and plot in the Presentation   and eventually convey this to others.}

\subsection{Who may this Presentation Review + EN be useful to?}

{\small In order to avoid disappointments,  it  is worth to stress that this Review is neither  designed as a public outreach talk  on Dark Matter, nor as a Seminar  that advertises a hot topic in Cosmology.

 The target of this Presentation Review is  the "Community" of  thousands of  physicists and astrophysicists whose work   already touches, in many different ways, the subject  of dark matter in galaxies and that  would like  to become more  aware of  the  phenomenology behind this subject,  possibly   relevant  for their research. 

The success of DMAW 2010 has shown that  this community is wide and that our dedicated   efforts are  well received.
We hope that many people will be interested in  downloading this Presentation + the Explanatory notes:

i) People who would like to follow this innovative approach so as to learn more about this increasingly important subject in Cosmology, Extragalactic Astrophysics and Astroparticle Physics, which they may often meet with in their own research work.

ii) PhD Students in any field of Astrophysics and Astroparticles

iii) Newcomers in Cosmology from different research fields.

iv) People who would like to deliver a well-tested Presentation Review  to colleagues or students, for the reasons given above. 

v) People that would like to  have a reliable Review Presentation on an important  subject  that may be of  help  in the preparation of  talks or lectures, centred either on this or on other adjacent topics.}

\subsection{Downloading}
\small{

The Presentation Review, the Explanatory Notes, the two movies that run inside the Presentation and the  Comments received for the  DMAW 2010 Initiative can  be downloaded at: 

\url{http://www.sissa.it/ap/dmg/dmaw_presentation.html} 

In detail,  the  names of the files of the downloadable  material  are:

The Presentation Review Talk  "Dark Matter in Galaxies":  DMGAL2010.pptx

   Explanatory Notes:  EXPLANOTES.pdf    

 Movies in the Presentation:  movie1.mp4, movie2.wmv 

The DMAW2010  Speakers' Reports and Comments:  Comments.pdf

}

\subsection{How to use this Presentation Review + EN ?}

{\small The following may be obvious but, to avoid misunderstandings,  must be stressed  for a correct and successful use of this material. The use of this  innovative product requires significantly  more time and personal  effort than reading  a    Review Paper  published in a Journal or in a Conference Proceedings Book. Users with profile (i-iii) above will need to invest and work  a couple of half- days to  read  the Notes and to  play the Presentation.  For users with profile (iv) the requested  effort is obviously greater. For users with profile (v), instead, the material we provide can save their working time and enable them to get improved results with less effort. Finally, let us talk about the users$^2$, that is,  those who will be  in the audiences of the users with profile (iv). They are the best off: with  90 minutes of attention they will get a good knowledge of the field, as it  was the case for the audiences of DMAW 2010.
}

\subsection{How to refer to/cite  this Presentation Review + EN ?}

{\small Differently from the usual situation   for reviews  published in Journals and Books,  the present  new scientific product should not be  cited  in papers  as the {\it unique}  support reference for a  result or  for a  view, also if they are  its core subject. We believe that the credit to the  results that this  review wants to  diffuse should go to the original papers.  On the other hand,  we would much like  to ask  people  to   advertise this Presentation Review  as much as possible in  papers  and in other occasions (e.g. Social Networks etc), in terms of the aims described in Section 1.2}

\subsection{ Comments and criticisms}

They are  welcomed,  especially if they consist  in  modifications and additions  in  a form  that,  in case, allow us   easily  to take them into account. 
   
\subsection{Acknowledgments}  We warmly and sincerely  thank our colleagues, all  the speakers, the LOC and the  SOC because,  without their willingness, suggestions and support, this product so as the whole DMAW 2010 initiative  would not have been realized.

 cfrigerio@ufabc.edu.br; Andrea.Lapi@roma2.infn.it, salucci@sissa.it

\newpage

\section{ The Review Talk: Dark Matter in Galaxies  }

 \subsection*{Abstract} 

This Review Talk concerns, in a  detailed way,
the mass discrepancy  phenomenon detected  in galaxies that usually we  account by postulating  
the presence of a non luminous non baryonic component. In the theoretical framework of Newtonian Gravity and Dark Matter Halos,    we start by recalling  the properties of the latter,    as emerging   from the  state-of-the-art of  the numerical simulations performed in the current $\Lambda CDM$ scenario of cosmological  structure formation and evolution. We then report the simple,  but  much-telling,   phenomenology of the distribution of dark and luminous  matter in Spirals, Ellipticals, and dwarf
Spheroidals.  It will be  shown  that   a coherent  observational framework  emerges from reliable data of different  large samples of objects.  The findings come after  applying different methods of investigation to different tracers of the gravitational field. They include RCs  and dispersion velocities profiles fitting, X-ray emitting  gas properties analysis, weak and strong lens signal mass decompositions, analysis of  halo and baryonic  mass functions.  We will then highlight  the  impressive evidence that the distribution of dark and luminous matter are closely correlated and that have  universal characteristics.  Hints on how this  phenomenological  scenario of the mass distribution in galaxies,  including the Milky Way   and the   nearby ones, has  a cosmological role,   are given.  Finally,  we discuss the constraints on the  elusive nature of the dark matter particle   that observations pose to  its direct and indirect searches. 

\newpage

\section{Explanatory Notes for the Slides}

\section*{Introduction} 
 
\subsection*{Slide 2}   
Dark Matter is a main actor  in Cosmology. It rules the processes of formation and evolution of all  the  structures  of the Universe, of which  today it constitutes   the great majority of the mass and  an important fraction of the mass energy. Furthermore, it  is  likely made of an elementary particle, beyond the Standard Model, but yet not detected.  However, this review will focus on just  one specific aspect of this mysterious component,  although a very crucial one:  its distribution in  galaxies, examined  also in relation with that of the ordinary matter.   The phenomenological  scenario  discussed in this  Review Seminar  can result a precious and  unique way to   understand how dark halos and their galaxies  formed and what they are really made of and maybe approach ourselves to new laws or processes of Nature.

\subsection*{Slides 4}

Galaxies, i.e. Spirals, Ellipticals, and Dwarfs spheroidals, have very likely just one type of dark halo. Their luminous components, instead, show a striking variety  in  morphology  and in the values of  their structural  physical quantities.

The range in magnitudes and central surface densities  is 15 mag and 16
mag/arcsec$^2$.

The distribution of the  luminous matter is given by a
stellar disk + a  spheroidal central bulge and an extended HI disk in
spirals and  by a stellar spheroid in ellipticals and dSphs.

\subsection*{Slide 5}
The best physical way to introduce the ``phenomenon'' of Dark Matter is the following:
let us define $M(r)$ the mass distribution of the gravitating matter and
$M_L(r)$ that of all baryonic components.
We can obtain  both distributions from observations;
we first realize that, as we start moving from a certain radius $ r_T$,
function of galaxy luminosity and Hubble Type, outwards, these distributions increasingly do not match, $d\log M/d \log r >d \log M_L/d \log r$.

We are then forced to introduce a non luminous component whose mass
profile $M_H(r)$ accounts for the disagreement:

$$
\frac{d \log M(r)}{d \log r} = \frac{M_L(r)}{M(r)} \frac{d \log M_L}{d \log r} + \frac{M_H(r)}{M(r)} \frac{d \log M_H}{d \log r}.
$$

This immediately shows that the phenomenon of the mass discrepancy in
galaxies emerges  from the discordant value of the  radial derivative of the
mass distribution, more precisely, from that of the circular velocity
or that of the dispersion velocity, two quantities that measure it:
$M \propto r\  V^2(r) \propto \sigma^2(r)$.

The DM phenomenon can be investigated only if we know
well the distribution of luminous matter and we can accurately
measure the distribution of the gravitating matter.

\subsection*{Slides 6}

We briefly  present the aspects related to the  current theory of galaxy formation  that are   mostly relevant 
for this core topic of this Review.

\subsection*{Slides 7-8}

It is worthwhile to briefly  recall the predictions for the structure of DM
halos coming from $\Lambda$CDM, the current theory of cosmological structure
formation  which is in well agreement with a number of cosmological observations.

A fundamental prediction, emerging from N-body
simulations performed in this scenario is that
virialized dark matter halos have a universal spherically averaged
density profile, $\rho_H(r)= \rho_{CDM} (r)$ \citep*{1997ApJ...490..493N}:
$$
\rho_{CDM}(r) = \frac{\rho_s}{(r/r_s)(1+r/r_s)^2},
$$
where $\rho_s$ and $r_s$ are strongly correlated
\citep{2002ApJ...568...52W,  2006ApJ...652...71W}, $r_{\rm s} \simeq 8.8
\left( \frac{M_{vir}}{10^{11}{\rm M}_{\odot}} \right)^{0.46} {\rm
kpc}$. The concentration  parameter $c\equiv r_s/R_{vir}$ is found to be a weak function of
mass \citep{2010arXiv1002.3660K} but a  crucial quantity  in determining 
the density shape at intermediate radii.

Notice that for any halo mass distribution $M_H(r)$:  $M_{vir}\equiv M_H(R_{vir})$

The recent Aquarius simulation, the highest resolution to date, renders a single dark  matter halo using 4.4 billion particles, of which 1.1 billion within the virial radius. 
We show spherically averaged density (left) and circular velocity (right) profiles.
Curves with different colours correspond to different resolution runs, and the corresponding  resolution limits are highlighted with arrows. 
These show that  the spherically-averaged density profiles of $\Lambda$CDM haloes deviate  slightly but systematically from the form found  by NFW by using  a much lesser number of particles and become progressively shallower inwards.   However,  these resolution differences are quite irrelevant and   the  NFW circular velocity well represent $\Lambda$CDM halos. Notice that  the latter  is almost never flat.  

In any case,  the actual profile  is of very  uncertain origin:   departures from  NFW shape are also seen in the  halo  velocity dispersion profile $\sigma$ and  the pseudo phase-space density emerges as a power law of radius: $\rho/ \sigma^3 \propto r^{1.8}$  e.g.   \citep{2009ApJ...692..174L}

\subsection*{Slide 9}

  The movie shows  the formation of one of the Aquarius halos, over nearly the full age of the Universe. The camera position moves slowly around the forming galactic halo, pointing towards its centre at all times.

\section{Dark Matter in Spirals}

\subsection*{Slide 11}

The stars are distributed in  a  thin disk with surface luminosity
\citep{1970ApJ...160..811F}:
$$
I(R) ~=I_0 ~ e^{-R/R_D}=\frac{M_{D}}{2 \pi R_{D}^{2}}\: e^{-R/R_{D}}
$$
where $R_D$  is the disk length scale,  $I_0$  is the central value and $M_D$ is the disk mass.
The light profile does not depend
on galaxy  luminosity and  the disk length-scale $R_D$ sets a consistent reference  scale in all objects and the extent of the
distribution of the luminous component,  we take  $R_{opt}=3.2 R_D$ as  the stellar disk size. 

 Although the  mass modeling of some individual object and certain
investigations require  to accurately deal  with the occasional presence of  non exponential stellar
disks  and/or with the  quite common  presence of a central bulge,  the above equation
well represents, for the aim
of this talk, the typical  distribution of stars in spirals 

\subsection*{Slide 12}

Spirals have a (HI + He) gaseous disk with a (very) slowing decreasing  surface density.
This disk,  at outer radii $R>R_{opt}$,  especially  in the case of low
luminosity objects,  is the major baryonic component and therefore must be
carefully considered in the mass modeling. An inner  H$_2$ disk is also present but  it is
negligible with respect to the stellar one \citep{2002ApJ...569..157W}

\subsection*{Slide 13}

The kinematics in spirals (and more rarely in other objects)  is measured by the  Doppler Shift  of  well-known  
emission lines of  particular  tracers of the gravitational potential: HI, CO and H$_\alpha$.

 RadioTelescospes  measure  the galaxy kinematics  by exploiting the   $21 \  cm$ emission line. Notice the  importance of  the spatial  resolution:  since if we want to map a galaxy  with a sufficient number of independent data also  in the inner regions,  the  HI emission disk size must be  at least 15-20 times bigger than  the radiotelescope resolution beam,  i.e. $\sim \  200  '' $. There are relatively  few objects in the  Universe complying with this requirement.  

 WSRT (Westerbork Synthesis Radio Telescope) has a resolution of $12\ ''$,  VLA (Very Large Array) of  $5 '' - 15 '' $,  ATCA (Australia Telescope Compact Array) of  $15 ''$. The  IRAM interferometer  at Plateau de Bure  measures  CO RCs by means of the CO(1-0) transition at 2.6 mm, at s spatial resolution of  $ 3 ''$.

\subsection*{Slide 14}

Gran Telescopio Canarias  is a 10.4 m diameter reflecting telescope at the Roque de los Muchachos Observatory on the island of La Palma.

In the  summit of Hawaii  dormant Mauna Kea volcano there  are the twin Keck telescopes, the optical and the  infrared.
The telescopes primary mirrors are 10 meters in diameter and are each composed of 36 hexagonal
segments that work in concert as a single piece of reflective glass.

The Large Binocular Telescope  is located on 3,300 m  Mount Graham in  Arizona. The LBT is one of the world's highest resolution and most technologically advanced optical telescopes;
the combined aperture of its two mirrors makes it the largest optical telescope in the world.

On Paranal ESO operates the ESO Very Large Telescope (VLT) with four 8.2-m telescopes. Each of them  provides one Cassegrain and two Nasmyth focus stations for facility instruments.

In measuring optical RCs, the class of the telescope is not always decisive, but of course important. 
Telescopes in  the  3-4 meter  class are sufficient to provide high-quality optical kinematics 
for local objects. The biggest telescopes  are necessary to measure the RCs at high redshifts.

Instead,  in  measuring dispersion velocities, new science  usually requires  the best telescopes,
also for local objects.

\subsection*{Slides 15-16}

We measure recessional velocities   by Doppler shift, and from these, the   rotation of the  disk.  In the process,
  we obtain the sky coordinates of the galaxy kinematical center, its systemic velocity, the degree of symmetry,
the  inclination angle and the circular velocity $V(R)$.  Optical
high quality kinematics  can   obtain circular velocity at a resolution  $ < R_D/10$.

\subsection*{Slides 17}

During the 70's there was running the idea of a  discrepancy between kinematics
and photometry,  already hinted in \citet{1970ApJ...160..811F} remarks about the NGC 300 RC.
A  topical moment was when  Rubin published 20  optical RCs extended to $R_{opt}$
that were still rising or flattish at the last measured point \citep{1980ApJ...239...50R}. As a guide, we show the profiles of the stellar  disk contribution if this  was   the sole contribution to the circular velocity.

Let us notice, however, that the actual  contribution due to the stars,  in view of the presence of a  central bulge and, in some case,  of a not exponential disk,  may be   inside $R_{opt} $   flatter than how it has been drawn (but never rising).

\subsection{Slide 18} 

The decisive measures that detected a dark component around spirals without doubt were   obtained  from the  21-cm RCs.
Notice that  HI line require radio telescopes with adequate sensitivity to detect the weak signals, and sufficient angular resolution to resolve the details. Only in the course of several decades the
equipment improved,  to arrive  in the late 70' at conclusive results.  A further
improvement occurred later  by using  interferometers, based on combining the signals of two or more telescopes,
and, by making use of the principle of earth rotation synthesis.
Today maps of the integrated HI distribution and the radial velocity field  are at a resolution of $< \ 10"$.

\citet{1978PhDT.......195B, 1981AJ.....86.1791B, 1981AJ.....86.1825B}  collected data and compiled HI RCs for 25
galaxies well extended beyond the optical radius $R_{opt}$,  indicated by a red bar on each curve.
 One obvious case is  NGC 5055 shown  in the slide.
 It is immediate that the  RCs do not decline in a Keplerian way $\propto R^{-1/2}$ beyond this radius.

More quantitatively  \citet{1979A&A....79..281B} obtained the surface
photometry for these objects  and found, in their outer
parts, a clear increase with radius of the {\it local } mass-to-light
ratios $\frac {dM/dR} {dL/dR}$.
If light traces the gravitating mass,
this quantity is expected to be constant with radius, equal to the
average mass -to- light ratio of the stellar disk.

\subsection*{Slide 19}

The  radial Tully-Fisher relationship gives a most straightforward
evidence for dark matter in spirals. The original paper of the standard TF relation is:  \citet{1977A&A....54..661T}.  

At different galactocentric distances,  measured in units of the optical size, $R_i\equiv  i \   R_{opt}$ ($i=0.2, ..., 1$), it has been observed the existence of  a family
of independent Tully-Fisher-like relationships, $M_{band} = b_i + a_i
\ log\  V(R_i)$ with $M_{band}$,  the magnitude in a specific band, in detail  the   (R, I) bands.

This ensemble of  relations,  we call the Radial Tully-Fisher  \citep{2007MNRAS.377..507Y},
constraints  the mass distribution in these objects.
It shows  a large systematic variation  of  the slopes $a_i$ with $R_i $:
they range across the disk  between $-4$ and $-8$. Moreover the RTF has a very small r.m.s. scatter at any radius.
This variation,  coupled with the smallness of the scatter,   rules out the
case in which the light follows the gravitating mass,  shown in the figure as the  black line predicting:  $M_I= const_i  + 10 \ log\  V(R_i)$ for  the case of a constant stellar mass-to- light ratio. However,  also in the case of a  magnitude dependence of the latter quantity, the slopes  of the RTF could be different from  the value of 10,  but they should  be  all equal,  at every   $R_i$. 

The RTF rules out mass models with no DM  or with the same dark-to-luminous-mass fraction within $i\ R_{opt}$ in any galaxy.

\subsection*{Slides 20-21}

Looking to individual RCs  (right) we realize that their profiles are a
function of the galaxy magnitude, a feature that points towards an 
universality of the spiral kinematics.
\citet*{1996MNRAS.281...27P} built 9
coadded RCs from about 700 individual RCs.
The whole $I$-band luminosity $-16.3 < {\rm M}_I< -23.4$ of spirals was divided in 9 luminosity intervals,
each of them with $\sim 1500$ velocity measurements.
For each interval the corresponding data were coadded in radial bins of
size $0.3 \,R_D $ to build the synthetic curves $V_{rot}(R/R_{opt},
M_I)$ out to $\sim 1.2 R_{opt}$,  $R_{opt}=3.2  \ R_D$.

The  coadded  RCs result  free from
observational errors and non-axisymmetric disturbances
present in individual RCs see Fig 1 and B1  of \citet*{1996MNRAS.281...27P}   and show a very small r.m.s.
This result was confirmed by a similar analysis performed with 3000 RCs by \citet*{2006ApJ...640..751C} (points in the plot).

\subsection*{Slides 22}

The URC is built from observations out to 6 $R_D$. There are not reliable kinematical tracers outside this radius.
It is possible, however,  to extrapolate it  out to virial radius (see \citet{2007MNRAS.378...41S}), the bonafide radius of  DM halos,  (related to  the virial mass by $R_{vir}=260 (M_{vir}/(10^{12} M_\odot))^{1/3}$
kpc)  by using the galaxy virial mass $M_{vir}$ and the virial velocity $V_{vir}^2= G\ M_{vir}/R_{vir}$. 
The former can be  obtained  by weak lensing  method \citep{2006MNRAS.368..715M, 2009MNRAS.398..635M} (left plot in the slide),
and by correlating the galaxy  baryonic  mass function $dN/dM_{bar}$ with the theoretical DM halo mass function
$dN/dM_{vir}$ (right plot). Two different recent works that have applied this method  well agree with themselves and with weak lensing result.   \citep{2006ApJ...643...14S,2004MNRAS.353..189V,
2002ApJ...569..101M,2010ApJ...710..903M,2006ApJ...647..201C}.  Then, at least in first approximation: 
$$
 M_D= 2.3 \times  10^{10} \  \frac {(M_{vir}/M_s)^{3.1}} {(1 + (M_{vir}/M_s)^{2.2})}\ M_\odot
$$
 with $M_s= 3 \times 10^{11}$ that gives $V_{vir}$ in terms of $M_D$,  derived  by the (known)  inner URC.

\subsection*{Slides 23}

These results lead to the concept of the Universal Rotation Curve (URC) of spirals.
Let us define $x\equiv R/R_D$ and $L$ the galaxy
luminosity.
We find that,   for any $x$ and $L$, the Cosmic
Variance of $V(x,L)$, i.e. the variance of  the circular velocity    at a  same radius $x$ in galaxies of same luminosity $L$,
  is negligible compared to the large 
differences of  $V(x,L)$  i) in a  galaxy of luminosity $L$  when  $x$ varies  and ii)   at the  same radius $x$ when $L$ varies.

 Therefore, the circular velocity at a given radius, this
radius and the galaxy luminosity, lie on a smooth surface.

The URC is shown in the provided  movie2 that should be linked to this slide.

\subsection*{Slides 24}

The individual circular velocities $V(r)$,  so as the coadded curves  are
the equilibrium  circular velocities implied by the galaxy mass
distributions. This allow us to decompose them  in different mass
components contributions.
In detail,  the gravitational potentials of a spherical
stellar bulge, a DM halo, a stellar disk, a gaseous disk
$\phi_{tot}=\phi_b+ \phi_{H}+\phi_{disk}+\phi_{HI}$ lead to:
$$
V^2_{tot}(r)=r\frac{d}{dr}\phi_{tot}=V^2_b +
V^2_{H}+V^2_{disk}+V^2_{HI},
$$
with the Poisson equation relating the surface (spatial) densities to
the corresponding gravitational potentials.
In general, we  can neglect the contribution  of the bulge, but this can play a crucial role in certain cases.
HI surface density  yields $V^2_{gas}$, usually computed numerically.
The contribution from the stellar exponential disk is

$$ 
V_{disk}^{2}(r)=\frac{G M_{D}}{2R_{D}}
x^{2}B\left(\frac{x}{2}\right),
$$
where $x\equiv r/R_{D}$, $G$ and $B=I_{0}K_{0}-I_{1}K_{1}$, a
combination of Bessel functions.

For the DM component we follow the approach of (a  quite general) Burkert empirical profile \citep{1995ApJ...447L..25B}
$$
\rho (r)={\rho_0\, r_0^3 \over (r+r_0)\,(r^2+r_0^2)} 
$$
$r_0$ is the core radius and $\rho_0$ the central density.  $V^2_{halo}= G
M_{H}/r$, $M_{H}= \int_0^R  4 \pi r^2 \rho(r) dr$.

It is sometime  used the pseudo-isothermal density $
\rho (r)=\rho_0\, \frac {r_0^2} {(r^2+r_0^2)} $. We warn about  this often used  cored distribution (outside  the baryonic  
regions of galaxies where any cored distribution is equivalent).  In fact, it implies  a divergent  halo mass   
and a  {\it outer} density  profile  inconsistent with that of NFW, in disagreement with  observations. 

Finally,   we can  adopt   the NFW profile emerging out of N-body simulations.

The differences  among the different halo profiles are shown  in the plot.

The mass model has three free parameters: the disk mass, and two quantities
related to the DM (the halo central density and core radius for
Burkert halos). These are obtained by best fitting the data.

Notice that also for  different Hubble type  ellipticals, dSphs and for
different techniques (weak lensing, dynamical modelling of dispersion
velocity) the mass modelling  always reduces in fitting  the total gravitating mass  profile $M(r)$ by means of a  mass model that includes  two components:
$M_{L}(P_1,r)$ and $M_{H}(P_1,P_2,r)$ and it has,  usually,  three free parameters $P_i$  (the mass in stars, the halo mass, the halo length scale).

\subsection*{Slides 25}

We start by  mass modelling the URC.

A Burkert halo + Freeman disk perfectly reproduce the  coadded RCs at any luminosity.
The luminous regions $R< 2 R_D$  of the lowest  luminosity objects are dark matter
dominated, while in  those  of high luminosity objects,  the stellar disk is the main  mass component.

In any object,  the dark component increases monotonically its importance on the circular velocity  as $ R$ increases.

Importantly, the
structural DM and LM parameters are related among themselves and with
luminosity.
In the 3D figure we see that spiral galaxies in the 4D
space defined by  central DM density, core radius, luminosity, fraction of DM at $R_{opt}$ lie on a curve.
On a physical side,  we realize that
smaller galaxies are denser and have a higher proportion of dark matter.

\subsection*{Slides 26-27}

The relationships between halo structural quantities and luminosity
must be investigated also via mass modelling of individual galaxies \citep{2004IAUS..220..377K}
that sometime is helped by assuming that the stars dominate the very
innermost regions.
The results confirm the picture derived from the
URC and point to the quantity $\rho_0 r_0$ which turns out to be  constant
in galaxies. Cored halos seem indispensable and smaller objects have  always more proportion of dark matter.

\subsection*{Slides 28-29  }

The investigation of the  density  distribution of DM  around spirals  to
check whether it  complies with the `raw'  $\Lambda$CDM predictions needs
careful model independent analysis of individual RCs.

This has been done  by using  suitable objects \citep {2005ApJ...634L.145G}, and also large  galaxy samples, including the recent The HI Nearby Galaxy Survey of uniform and high quality data, with a significant number of rotation curves in which non-circular motions are small.

The available rotation curves are successfully  fitted by a cored-halo + the  stellar/HI disks.  For  DDO 47 the separate  contributions   are indicated.   Instead,  they are badly  fitted  by the  NFW  (or similar) halos  + stellar/HI disks model,  of which for IC 2575 we plot the separate components.  The blue region indicates the discrepancy between NFW + disks models and actual data.

The results in  these galaxies are quite typical. In addition to poor RC fits the  latter mass  model  often leads to  very large virial halo masses, and  stellar mass-to-light ratio that result too low  for the  observed luminosities and colors of the  galaxies.  
It is also proved that tri-axiality and non-circular motions  cannot explain such  CDM/NFW cusp discrepancy  with the observed RC data of spirals and LSBs
\citep{2004MNRAS.351..903G,2008AJ....136.2648D,2008ApJ...676..920K,2008AJ....136.2761O,2008MNRAS.383..297S,2008AJ....136.2720T,2009MNRAS.397.1169D}.

It is however worth recalling  that several physical processes that modify the original cosmological  NFW halo  profiles to meet with the observed ones  have been proposed.  This  discussion is beyond the aims of this  talk,  especially because consensus is far from being reached about the  net effect of   (baryonic) physical processes  on the dark matter distribution: some physical processes make it close to the observed (cored) \citep{2010Natur.463..203G}, while others make the NFW cusp even steeper, e.g.  \citep{2004ApJ...616...16G}.

\section{Dark matter in ellipticals}

\subsection*{Slide 32}

In Ellipticals the surface brightness of stellar spheroid, the main baryonic component,  follows
a S\'ersic (de Vaucouleurs) law:
$$
I(R)= I_0 \ {\rm dex} [-b_n (R/R_e)^{1/n}-1]
$$
where $n$ is the index of S\'ersic defining the degree of concentration of the light and $R_e$ is the half light radius.
For $n=4$ we obtain the well-known de Vaucoulers profile 
By deprojecting the surface density $I(R)$  we obtain the luminosity density $j(r)$ and by assuming a radially constant stellar mass to light ratio $(M/L)_\star $,  the spheroid density $\rho_{sph} (r) $.
The central surface brightness $\mu_0$ is  given by:  $\mu_0=2.5 \log I_0 +const$.

\subsection*{Slide 33}

To derive the gravitational potential in Ellipticals and from this to infer the  dark matter distribution is significantly more complicated than in  spirals.
The former  can be obtained, always caveat a number of crucial  assumptions,  from dispersion velocities of stars or planetary Nebulae, from  the X-ray properties of the emitting hot gas halos that lies around them, or from a  combination of weak and strong lensing data.  

 Jeans equation is a crucial link between observations and data.   We infer from the  observed motions the underlying gravitational potential.  The knowledge of the distribution of dark and luminous matter in these systems of different morphology, age and formation process,  is however indispensable.

\subsection*{Slides 34 }

In ellipticals the kinematics is complex,   the stars are in gravitational
equilibrium  by   balancing  the
gravitational potential they are subject to with 3D motions, whose r.m.s
exerts a pressure. Obviously,  from the motions of stars in a galaxy we cannot measure the radial/tangential velocity dispersions, directly linked to the mass profile,  but recessional velocities or  projected and    aperture velocity dispersions, quantities that are a  much more indirect tracer of the gravitational
potential, as shown by the formula in the slide, valid  for  case of isotropic orbits. The kinematics and its analysis  becomes  more complicated when this assumption is relaxed  and/or some rotation is present.

In this case the full 2D kinematics is absolutely required and we must solve (with some assumption) the anisotropic Jeans equation, i.e. that involves  also the higher moments of the LVDS at any position in the galaxy \citep{2008MNRAS.390...71C,2005MNRAS.357.1113K}.

\subsection*{Slide 35}

Ellipticals  are quite compact objects,  with respect to spirals. Stellar kinematics of the spheroid traces a very inner region of the DM halo, where likely it has small dynamical effect. Moreover, the analysis is far from straightforward.  This can be
seen when we try to reproduce  the (observed) velocity dispersion profile $\sigma_{ap}(R) $
with a  reasonable model made of a dark halo and a luminous mass component. We take a galaxy with a spheroid
of   $10^{11} M_\odot$ with the standard  S\'ersic parameter $n=3$,   a NFW halo with  $c=7$ 
and  a mass 20 times bigger that of  the spheroid.  In this mock  object the DM has a an  dynamical important role, and it would emerge  if its gravitational potential was probed by  the rotation curve. However, when probed by  $\sigma_{ap} $, the nature itself of this kinematical hides the dynamical effect of the dark component \citep{2005MNRAS.362...95M}. For a cored halo or  with the presence of anisotropy the situation would be worse. 

A similar result  comes from  the work of \citet{2010MNRAS.tmp.1737T} in which it is  shown  that an assumed typical NFW  DM halo around an elliptical, i.e. with  
$M_{vir}<10^{14} \  M_\odot$,   contributes
to the observed  central  aperture  velocity dispersion (e.g. to  $ \sigma_{ap}(1/2 \ R_e)$) in a negligible way:   
$ \sigma_{ap,H}(1/2 \ R_e)^2 < (100~ {\rm km/s})^2  <<
(\sigma_{ap})^2$. For cored halos the contribution is even smaller. Only
the smallest ellipticals and dwarfs spheroidals in which $\sigma_{ap}(1/2 \
R_e) < 100 \  km/s $ , the stellar velocity dispersion may probe the dark component.

\subsection*{Slide 36}

Important information on the mass distribution can be obtained from
the  Fundamental Plane.
For  virialized stable objects  one expects 
the balance between  the potential and kinetic energies.
$$
\sigma^2 \propto \frac{GM_{dyn}}{R}
       \propto \frac{M_{dyn}}{L}\,\frac{L}{R^2}\, R
       \propto \frac{M_{dyn}}{L}\,I\,R,
$$
where $\sigma$ is a velocity dispersion, $I$ is a surface luminosity, $R$ is a scale, and $M_{dyn}/L$ is the mass-to-light ratio.
This implies a relationship between the observed velocity dispersion
$\sigma_e^2$, $R_e$   and surface luminosity $I_e\equiv L_e/(2 \pi R_e^2)$.
i.e. a (Fundamental ) Plane in the space of these 3 quantities. 
From  the virial equilibrium,  we expect: $ R_e = \sigma_0^a/I_0^b $.

This relationship is found in ellipticals and S0:  in the 3D space of ($R_e, \sigma_0, <I_e>)$, where  the latter is  the surface brightness inside $R_e$ \citep{1996MNRAS.280..167J} we have: 
$$
log  \  R_e= 1.24 \ log \ \sigma_0 - 0.82 \ Log <I>_e
$$
the scatter is very small, 0.07 dex in $log \  R_e $

A similar results is found  in the   large sample of ellipticals that include 6000 objects of  the digitalized survey  SDSS
\citep{2009MNRAS.396.1171H}.

The  very small  scatter of objects around this plane suggests some  degree of universality in their {\it  inner}  mass profile. 

The value  of the coefficient $a $ is  different from that expected by
the virial theorem, $a=2$ in the case  in which ellipticals  have the same dynamical mass-to-light ratio. This tilt of the FP can  explained   in a dependence of the stellar mass
to light ratio  $(M_{sph}/L)$ on  luminosity (spheroid mass). In fact,  FP residuals are  found correlated with stellar
population characteristics as line- strength indices and  are anticorrelated with the abundance ratio 
$\alpha / Fe$, \citep{2009MNRAS.397...75G}.

\subsection*{Slide 37-39}

Planetary Nebulae (PN) provide us with the elliptical galaxy kinematics out to many effective radii, \citep{2009ApJ...691..228M}.  The use of dedicated instruments, like the Planetary Nebula Spectrograph  \citep{2002PASP..114.1234D} has strongly powered the technique,  so that distances  comparably to  those  reached by extended HI disks in spirals 
can be probed \citep{2009MNRAS.394.1249C}.

In each galaxy  ~200 individual PN radial velocity measurements are
usually obtained.   By binning PN radial velocities in elliptical
bins, or in strips over the major and minor axes  (see for details
\citep {2001A&A...377..784N}),  reliable  2D kinematical maps have
been derived \citep{2008AN....329..912C}. These maps  led to independent
and high quality projected velocity dispersion measurements   out to $~(5-8) R_{e}$,  with  a typical spatial resolution of $1/3 \  R_{e}$.

A systematic study of the velocity dispersions in  a dozen of  objects has  revealed that  in the most  luminous  objects the velocity dispersion  tends to  show a regular, monotonic   flat  profile, that poses no problem in the subsequent mass modelling,  other  than its uniqueness {\citep{2009MNRAS.394.1249C, 2010MNRAS.tmp.1835N}.

In  low luminosity objects, instead,     the PNs  velocity dispersion profile show a large  pseudo-and ultra-keplerian decline:  $\sigma_P \propto R^a$ with $a<-0.5$. To analyse these data  obviously requires  a full  modelling  of a very likely complex dynamics, that includes also an appropriate treatment of the orbital anisotropies of the tracers.

For objects with ``regular'' kinematics the best case analysed so far is  NGC 4374 \citep{2010MNRAS.tmp.1835N} where a flat dispersion profile  has been derived  from   about 450 PNes and then modeled  by means of Jeans analysis.

As result,  a variety of halo  model profiles including  NFW,    some of its variants  and the cored URC halo, in cooperation with a spheroid with typical  mass-to-light ratios,  well fit the data, confirming the presence  a very massive dark halo. Notice that  in this elliptical the kinetical energy $\propto  V_{rms} ^2 $ has both  rotational and random motions component: $V_{rms}=  (V_{rot}^2 + \sigma_P^2)^{1/2}$ .

\subsection*{Slides 40-42}

Gravity bends light and this can be used to weigh the galaxy mass, e.g.   the pioneering work by \citep{1996MNRAS.283..837S}.
We know that a  background round galaxy (of semi axis $a/b=1$) whose line of the sight passes at distance $R_P$ from a the center of a galaxy put
at a convenient distance (half way between us the  background)  will be
seen with a (very small)  ellipticity $a/b-1\simeq 10^{-3}$. 
The ellipticity of a galaxy image is an unbiased estimate of the local shear.
When averaged on a very large number of source galaxies  whose lines of sight  all pass around a lens, 
 we can  meaningfully measure the tangential shear $\gamma_t$ of the lens.
The lens equation relates $\gamma_t$ with the distribution of matter in the lensing
galaxy:

$$
\gamma_t= (\bar \Sigma- \Sigma(R))/\Sigma_c,
$$
where $\Sigma(R)=2 \int_{0}^{\infty} \rho(R,z)dz $ is the projected mass density of the object distorting the galaxy image,
at projected radius R and $\bar{\Sigma}(R)= \frac{2}{R^2} \int_{0}^{R} x \Sigma(x) dx$ is the mean projected mass density interior to the radius $R$.
$\Sigma_c\equiv \frac{c^2}{4 \pi G}  \frac{D_s}{D_l D_{ls}}$,
where $D_s$ and $D_l$ are the distances from the observer to the source and lens, respectively,
and $D_{ls}$ is the source-lens distance.
The above relations directly relate observed signals with the underlying DM halo density. 
A  very  detailed explanation of the weak lensing phenomenon can be found in the second chapter of \citet{2006glsw.book..269S}.

Let us stress  that we discuss this effect in the section of elliptical in that it has been applied mostly for these objects, however there are applications (even in this Review) for Spirals.

The DM distribution is obtained
by  fitting the observed shear with a chosen halo  density profile with 2 free parameters.
The method is clean because it is free from the baryonic contribution,
but the  signal is intrinsically  weak.
\citet{2009MNRAS.398..635M,2006MNRAS.368..715M} measured the shear
around  $1.7 \times 10^5 $ isolated galaxies of different luminosities and Hubble Type,  by using  $3 \times 10^7$ SDSS galaxies as  sources.  The measurements extended out to $500-1000$ kpc reaching out
the virial radius of the lens galaxies.  Although  the structural  DM parameters were determined with quite large  uncertainties,
the NFW and Burkert halo profiles, not much different at outer radii,
both agree with the data. No relevant  difference in the DM distribution around ellipticals or spirals emerged.

Very importantly, the virial halo  masses so determined are  found to correlate with the galaxy luminosities.  When  this  relation is
coupled with the well known  mass vs. light relationship of the stellar
component,  it leads to the fact  that the amount of baryons $M_{bar}$ today  in a galaxy  is  a
function of  $M_{vir}$, the   mass  of the halo  surrounding the galaxy. We consider this as  the main
 relationship  of the galaxy formation process. It is interesting  to note that  $M_{vir}/M_{bar}>> 7 $, the cosmological value.

Noticeably, when  we fit the tangential shear with a Burkert profile,  we obtain the  same values of  halo structural parameters $\rho_0 \ , r_0$ derived by mass modelling the URC or individual galaxies \citep{2009MNRAS.397.1169D}.

\subsection*{Slide 43-44 }

\citet{2006ApJ...649..599K} performed a joint strong lensing +
stellar-dynamical analysis of a sample of 15 massive early-type
galaxies selected from the SLACS Survey. Given the smallness of the
radial range investigated, the total mass density was assumed  a power
law without loss of generality.  The spheroid component,      was
represented for simplicity, as an Hernquist profile, very similar
inside $R_e$ to a S\'ersic one. Velocity dispersions at $R_e$  provided the  additional constraint to the mass distribution, that was investigated   through Jeans Equation.

They found, independently of the  DM density profile in the region under investigation, that   these massive early-type galaxies have remarkably the same  inner {\it total}   density profiles ($\rho_{tot}=  \rho_{H}+\rho_{sph}\propto r^{-2}$). The figure  shows the logarithmic density slope of SLACS lens galaxies as a function of (normalized) Einstein radius. Moreover, they found  that  the stellar spheroid  accounts  for most of the  total mass inside $R_e$. 

In both cases, given the smallness of the dark component inside $R_e$,
the small radial range,  the not negligible observational error and  the relevance of the assumptions taken,   it is not possible to constrain  the actual profile of the dark halo.

\subsection*{Slide 45}
Isolated Ellipticals  have a  X-ray emitting halo of regular
morphology, that,   extends   out to very large radii, providing us with  a reliable
mass distribution.
The gravitating mass inside a radius $r$,   $M(r)$  can
be derived from their X-ray flux by assuming that the emitting  gas is in
hydrostatic equilibrium.
The observed  X-ray surface brightness, gives,  by
means of the well known  $\beta$-model,  the  gas density distribution.
From the hot  gas density and temperature profiles 
\citep*{1984ApJ...286..186F},  we have:
$$
M(<r) = {kT_\mathrm{g}(r) r \over  { G \mu m_p} }\Big({d\log\
{n}\over{d\log\  r}} + {d\log
\ T_\mathrm{g}(r)\over{d\log \ r}}\Big)
$$
where $T_{\rm g}$ is the gas temperature at radius $r$, $\rho_g$ is the gas
density, $k$ is the Boltzmann's constant, $G$ is the gravitational constant,
$\mu $ is the mean molecular weight $\mu=0.62$ \citep{2006MNRAS.369L..42E},
and $m_p$ is the mass of the proton.

\citet{2009A&A...501..157N} analyzed a sample of 22 objects with XMM-Newton and
Chandra data out to 10 $R_e$.
They confirmed  the presence of DM in any object, but  this component was found   dynamically important
only outside $R_e$, where the mass-to-light ratio is found to start to increase.
Furthermore, there is some hint that, outside this radius,  the halo mass
increases with radii quite steeply suggesting a  cored DM halo
density, for which the dark mass increase rate between $r^3$ and $r^2$.

\subsection*{Slide 46}
The gravitating matter in form of stars $M_*$ is a crucial quantity.
It defines the efficiency of the process of forming this (luminous) 
component from the cosmological hydrogen.
Usually, it is obtained as result of the kinematical mass
modelling, but in some case its resulting value has a very large 
uncertainty. This, in turn complicate the full
investigation on the Dark Matter distribution n. Moreover,
in low luminosity and DM dominated objects, the kinematics cannot 
provide a reliable estimate of the stellar disk mass.

It is well known that the galaxy luminosity is related to its stellar content. 
Hence the straightforward approach to obtain the stellar mass of a galaxy is to model its luminosity in terms 
of age, metallicity, initial mass function. This modelling is provided by Stellar Population Synthesis models.
This method is presented  in this section since it has important applications in Ellipticals, but it is applied  also to  Spirals and dSphs. 

The observed Spectral Energy Distribution (SED) of a galaxy, or selected colour indices or absorption lines, is fitted with models calculated under different assumptions regarding the IMF, the star formation history, the metallicity, etc. The same modelling naturally provides the amount of stellar mass that is implied by the different fits (for a classical paper see \citet {1975MmSAI..46....3T}. In practice, degeneracies between age, metallicities and dust content make it sometimes hard to select the actual physical model for the galaxy light just in terms of fit goodness. Observations are decisive to solve these intrinsic uncertainties.

While the best approach is obviously to fit the whole spectral energy distribution, as this allows a better handle of elusive degeneracies, often not all observational bands are available for galaxies. A simplified approach was suggested by
\citet{2001ApJ...550..212B} who noted that rather simple relationships exist between mass-to-light ratios in certain bands and some colour indices of population synthesis models (from various codes). These relations have been calibrated using spiral galaxies.
In the plot there are two different  $M/L$'s as    a function of (B-R), for a sequence of exponentially
declining star formation rate models of age 12 Gyr obtained by using a 
variety of Stellar Population Synthesis models. The red end of the lines represent 
a short burst of star formation,
the blue end represents a constant star formation rate model. The 
different models used are:
Bruzual \& Charlot, Kodama \& Arimoto, Schulz et
al. and PEGASE by Fioc \& Rocca-Volmerang all with a Salpeter Initial 
Mass Function.

Recent work considers in great depth the TP-AGB phase of stellar 
evolution while building the galaxy SED \citep{1998MNRAS.300..872M,  2005MNRAS.362..799M}
which were shown to provide a better fit to the spectral energy distribution of high-redshift galaxies
\citep{2006ApJ...652...85M}, allowing to better determine their stellar masses.
The resulting mass-to-light ratios, as a function 
of galaxy color are shown in the slide.

\subsection*{Slide 47}

 The photometric estimates of stellar disk/spheroid masses are tested 
with respect to the kinematical
estimate of the mass in stars.

In \citet{2007MNRAS.378...41S} the disk masses of 18 spiral galaxies of 
different luminosity and Hubble Type have been derived both by mass 
modelling their rotation curves and by fitting their SED with 
spectro-photometric models. The two estimates result in perfect 
proportionality and the agreement between the two different estimates is 
less than 0.15 dex.

In Elliptical galaxies the determination of dynamical mass inside $R_e$ 
can be performed
by assuming that a De Vaucoulers spheroid is the main component then:
$M_{dyn} = 1.5 \times 10^{10} (\sigma_e/200$ km/s$)^2 (R_e /kpc) M\_odot$
This dynamical estimate of the ``stellar'' mass is found to well 
correlate with the red
luminosity, i.e. with a measure of the photometric mass estimate 
\citep{2009MNRAS.396L..76S}.

\citet{2009A&A...501..461G} studied a sample of Ellipticals with 
Einstein rings and derived the total projected mass enclosed within them. Then, by using the 
SDSS multicolor photometry fitted the spectral energy distributions 
(SEDs) (ugriz magnitudes) with a three-parameter grid (age, 
star-formation timescale, and photometric mass) of Bruzual \& Charlot's 
and Maraston's composite stellar-population models, and various initial 
mass functions (IMFs). The two different mass estimates agreed with 0.2 dex.

A similar result has been obtained by studying a sample of ellipticals with detailed axisymmetric dynamical modelling of 2D kinematics  \citep{2006MNRAS.366.1126C}.  

Caution: the disagreement among dynamical and spectro-photometric 
estimates is about 0.15 dex, this value is small for cosmological 
implications, i.e. to build a stellar mass function from a luminosity 
function, but it is extremely large for mass modelling. The induced 
error, if we assume $(M_*/L) ^{SPS}$ in estimating the DM profile 

$$\rho_{H}(r)= \frac {1}{ 4 \pi r^2} d(M_{dyn}(r)-(M_*/L) ^{SPS} 
L(r))/dr
$$
can totally err the derived density profile.

\section{Dark matter in dwarf spheroidals}

\subsection*{Slide 50}

Dwarf spheroidal (dSph) galaxies are the smallest and least luminous galaxies in the Universe and so provide unique hints on the nature of DM and of the  galaxy formation process.
Early observations (e.g. \citet{1983ApJ...266L..11A}) of stellar kinematics in the brightest ($L_V \sim 10^{5-7}L_{V,\odot}$) of the Milky Way's dSph satellites found that they have central velocity dispersions of $\sim$ 10 km s$^{-1}$, larger than expected for self-gravitating, equilibrium stellar populations with scale radii of $\sim$ 100 pc.
The recent discoveries of ultra-faint Milky Way satellites
(e.g. \citet{2005AJ....129.2692W,2007ApJ...654..897B}) extend the range of dSph structural parameters by an order of magnitude in (luminous) scale radius and by three orders of magnitude in luminosity.

The distribution of total luminosity and of the  central stellar velocity dispersion in  star clusters and in dSphs overlap.
The faintest galaxies have approximately the same values of these physical parameters of star clusters, with galaxy luminosity extending down to $10^3 L_{\odot}$.
In  dSphs the  half-light radii, however, are significantly larger (hundreds of parsecs) than those of star clusters (tens of parsecs).
This leads, through the virial theorem, to significantly larger masses in dSph.

The values of central and global mass-to-light ratios for the gas-poor, low-luminosity, low surface brightness satellites  of the Milky Way are high, up to several hundred in solar units, making these systems the most dark matter dominated galaxies in the local Universe \citep{1998ARA&A..36..435M}.

\subsection*{Slide 51}

DSphs are old and apparently in dynamical equilibrium, with no dynamically-significant gas.
They contain a (small) number of stars, which provide  us with  collisionless tracer particles.
\citet{1983ApJ...266L..11A} measured the velocity dispersion profile of Draco based on observations of three carbon stars, finding a mass-to-light ratio an order of magnitude greater than that found for galactic globulars.
\citet{1997ASPC..116..259M} provided the first dispersion velocity profile of Fornax.
With the new generation of wide-field multi-object spectrographs on 4-8m telescopes, it is now viable to obtain sufficient high-quality kinematic data to determine  the gravitational potentials in all the Galactic satellites.

\subsection*{Slide 52}

\citet{2007ApJ...667L..53W} published high quality  velocity dispersion profiles for seven Milky Way satellites.
They are  generally flat out  to large radii, in terms of the half light radii  that determine the sizes of the baryonic matter in these objects. These data indicate that  almost all of  their  gravitating mass is unseen, i.e. in a DM halo. 

\subsection*{Slide 53-54}

The mass modelling of these object is not immediate, in that  the  distribution of the stars, that are the tracers of the gravitational fields,  must be known with precision. Moreover,   the relation between mass and dispersion velocity is not a  straightforward one:  in  a collisionless equilibrium system that of the stellar spheroids in dSph,  the Jeans equation  relates the kinematics of the tracer stellar population and the underlying (stellar plus dark) mass distribution. Assuming spherical symmetry, the mass profile can be derived as
$$
M(r)=-\frac{r^2}{G}\Big(\frac{1}{\nu} \frac{d\  \nu  \sigma^2_{r}}{dr}+ 2\frac{\beta \sigma^2_{r}}{r}\Big),
$$
where $\sigma_r(r)$, $\beta(r)$  and $\nu(r)$ are, respectively, the stellar velocity dispersion component radially toward the center of the mass distribution, the velocity anisotropy and the stellar density distribution.

Disadvantages in  modeling these systems are: i)  it is required  to take  some assumption on the velocity anisotropy  and  ii) derive the stellar density profile with precision, in order to  fit the observed velocity dispersion profile and  obtain a reliable  DM  distribution.

An advantage is that their  luminosities are so small that we can always exclude that baryons play a (complicated)  role in the mass modelling.

The surface brightness profiles are typically fit by a Plummer distribution \citep{1915MNRAS..76..107P}
$$
\nu(R)=\frac{\nu_0}{(1+R^2/R^2_b)^2},
$$
where  $R$ is the projected radius, $\nu_0$ is the central surface density and $R_b$ is stellar length-scale
Notice that,  if we want to derive the DM density profile (not its amount)  we must know with good accuracy the   distribution of the stellar component,  since
this quantity enters in the determination of the  mass {\it profile}  in a non-linear way. In Elliptical and Spirals, it is insread extremely rare that, given a set of kinematical data,  two different  stellar distributions but both  compatible with the measured  surface photometry,  yield, in the mass modelling,  different DM profiles.  

The analysis of the  mass distributions  obtained  from isotropic Jeans equation analysis of six dSphs by \citet{2007ApJ...663..948G} favor a cored dark matter density distribution.

However, \citet{2009ApJ..704.1274W}  applying  the Jeans equation to eight brightest dSphs and relaxing the isotropy assumption   found that  the dispersion velocities of these objects can be  fit by a  wide range of different  halo models and velocity anisotropies.
In particular cusped and cored dark matter mass models can fit the dispersion profiles equally well.

\subsection*{Slide 55 }

 A resolutive fact could be that  dSphs cored distribution structural parameters  surprisingly agree with those of Spirals and Ellipticals:
\citet{2009MNRAS.397.1169D}  have shown that a Burkert halo density profile reproduces the dispersion velocity  of six  Milky Way dSphs and  that the derived halo central densities and  core radii are consistent with the extrapolation of the relationship between these quantities found  in Spirals.
This intriguing result, beside supporting cored distribution in dSph,  could point to a common physical process responsible for the formation of cores in galactic halos of all sizes or to a strong coupling between the DM and luminous matter.

\subsection*{Slide 56 }

The dark+ stellar  mass within the effective  radii of  dSph galaxies
have been suspected for some years of showing a remarkably small range \citep{1998ARA&A..36..435M}. 
The available data  seem to be  consistent with an (apparent) minimum dark halo mass,  supported also  by  the presence of a strong relationship  between $M_*/L_V$ and   $L$     \citep{2006EAS....20..105W, 2007NuPhS.173...15G}.

More explicitly,  a relationship is found for which, 
within 300 pc, all object seem to possess the same mass,  of the order of $10^7 M_{\odot}$, 
\citep{1999ApJ...526..147C, 2008Natur.454.1096S,  2007ApJ...663..948G}. This relationship maybe be not intrinsic, but due to  the universality of the  profile of the density of the halos around galaxies. 

On the other hand,  if all  dSphs are embedded within 
an  "unique"   dark matter halo, this  could explain the finding  that their  masses increase as  power-law   in all objects, $M(r)\propto r^{1.4}$ \citep{2009ApJ...704.1274W}.

\subsection*{Slide 58}

DM halos around galaxies form a family of dark spheres with their virial mass being the order parameter.
It is possible to work out an  unified scheme for  their structural characteristics,  that,  let us remind,  are derived  by means of several different methods  in galaxies of different luminosities  and  morphologies.  From all available data a picture emerges:  DM halos  around spirals, LSBs and dSphs and Ellipticals can be well represented by a Burkert profile with a core radius $r_0$ and a central density $\rho_0$. 

Very intriguingly: the halo central surface density $ \propto \rho_0  \ r_0 \ $  is  found nearly constant and independent of galaxy luminosity. This result come from the mass models   of 1)  about 1000 coadded rotation curve of Spirals, 2)  individual dwarf irregulars and spiral galaxies of late and early types 3)  galaxy-galaxy weak lensing signals 4) kinematics of Local Group dwarf spheroidals.

It is found:  
$$
\ log (r_0 \ \rho_0) = 2.15 \pm  0.2$$
in units of $log (M_\odot /pc^2)$. 

This  results are obtained for galactic systems spanning over 14 magnitudes, belonging to different Hubble Types, and whose mass profiles have been determined by several independent methods. In the same objects, the constancy of $\rho_0 \ r_0$ is in sharp contrast to the systematical variations, by 3-5  orders of magnitude, of galaxy properties, including $\rho_0$ and central {\it stellar}  surface density. 

\subsection*{Slide 59}

A  connection between the  a-priori independent distributions  of  dark and luminous matter in galaxies emerges strongly. The velocity (mass)  profile, despite being the sum  in quadrature of {\it both}  components,  can surprisingly  be fully represented by a function that includes only one of them.  In fact, for Spirals, (but likely in galaxies of other Hubble Types) we have : $V(r)= F_{bar}(r/R_D, M_I)$ ($M_I $ is the I- band magnitude, left- side plot), but,  at the same time, we  also have:  $V(r) = F_{vir} (r/R_{vir}, M_{vir})$ (right-side plot)  \citep{2007MNRAS.378...41S}.

In this last case the  baryonic component breaks the  mass   invariance of the
RC profile due to  the  DM component, but in a mass dependent way. The differences of the inner RCs profile of high and low mass objects are  due to  differences in the fractional content of baryonic matter, but mysteriously, they, in turn,  depend on the halo mass.   

It may be worth recalling that  similar result (in qualitatively way) are present   in the  predictions of  $\Lambda$CDM, where  NFW halos  coupled with a stellar disks formed according to the  \citet{1998MNRAS.295..319M}    theory  leads to RCs Universal profile,  function of the halo virial mass.

\subsection*{Slide 60}

\citet {2010ApJ...710..903M} used a statistical approach   to determine the relationship between the stellar masses of galaxies and the masses of the dark matter halos in which they reside. They  obtained (following  \citet{2006ApJ...643...14S}) a  stellar-to-halo mass (SHM) relation by populating halos and subhalos in an N-body simulation with galaxies and requiring that the observed stellar mass function be reproduced. The existence of this non linear  relation is not  trivial: $M_{\star}$, the mass in stars,  does not scale proportionally to $M_{vir} $ and it is  much smaller than that expected by the Cosmological value (blu line) $\Omega_{baryons}/\Omega_{DM}$.

\citet{2010ApJ...717L..87W},  subtracted  in spirals  and  in  dSph  
the baryonic  contributions to  the overall  gravitational potential  to derive  the  contribution to  rotation curves
due to dark matter halos at intermediate radii. They found that, in physical units, these RCs  are  remarkably similar, with a mean curve given by

$$
\log  \ \frac{V_H} {(km/s)}   = 1.47^{+0.15}_{-0.19}  + 0.2 \ \log \ (r/kpc)
$$
 
Evaluation at specific radii of this Universal Profile immediately
generates two results  discussed above: a common mass for MW dSphs at fixed radius and a constant
DM central surface density for galaxies ranging from MW dSphs to spirals. This result is very remarkable: in  the  Universal Functions  discussed  above as the  URC and the  FP,  the radial unit  is always  measured in terms of the stellar  galaxy size  and not, as here, in physical units.

However, in dSphs it is possible to estimate ``the circular velocity'' only at one radius  in each  galaxy   \citep{2009ApJ...704.1274W},  namely  the half light radius, whose definition itself,however, may be not trivial \citep{2010ApJ...710..886W}.

This very recent result needs more investigation, though it  seems coherent with the phenomenology of the DM  distribution around galaxies.

\subsection*{Slide 61}

Let us define $R_e$ as  the effective radius encompassing {\it half}  of the mass of stellar component and  provides so a the measure of the size of the latter. At  $R_e $  in dispersion-supported
galaxies,  within a large range of luminosities, the  dynamical mass-to-light ($M/L$) ratios can be obtained \citep{2010MNRAS.406.1220W}.  
 At the high mass end, these estimates come from the   FP, while
at the low mass end, they come from Jeans modelling of dwarf spheroidal galaxies.

Confronting the emerging mass-to-light ratios, galaxies follow a characteristic U-shaped curve \citep{2010MNRAS.406.1220W},
 where $M/L$ is minimized for
$\sim L^*$ galaxies and rises at both low and high masses. This can be seen
in the plot, where the $I$-band $M/L$ ratio within the half-light radius is
plotted against the mass contained within the half-light radius.

One can interpret  this as the inverse efficiency with which galaxies retain and turn their initial
quotient of baryons into stars : $1/7 \  M_{vir} $. There must be two mechanisms that prevent galaxies
from turning their initial baryons into stars, one of which operates most efficiently
at low masses, switching on strongly below $M(R_e) < 10^8~M_{\odot}$, and one of
which operates most efficiently at high mass, gradually becoming more effective
at masses above $ M_(R_e) > 10^{11}~M_\odot$. Some suggested mechanisms for
the former include heating from re-ionization, supernova feedback, and tidal and
ram pressure stripping for those galaxies that are satellites, while  proposals 
for the latter mechanism are feedback from active galactic nuclei, virial heating
of gas in the deep potential well, and supernova feedback.

It is interesting that,  globular clusters, whose masses can be calculated using the same Jeans modelling
methods as those of dwarf spheroidal galaxies, are unambiguously not part of
this relation. They have roughly constant $M/L$ ratios, consistent with
that expected from their stellar populations.
Systems containing dark matter and those without dark matter are therefore
easily distinguished.

It is worth recalling that the same U shape is seen in {\it disk systems} at $R_{vir} $   \citep{2006ApJ...643...14S}  and it is shown as yellow region in the plot.

\section{Dark matter searches}

\subsection*{Slide 62}

To introduce the various candidates proposed as the Dark Particle and to  discuss  the present  status of their possible detection
 would need  a complete (long) Presentation Review. As matter of fact we  deal these subjects  in  a  brief section of our Review. This is certainly insufficient however  this Presentation is likely to be for some the first (serious) contact with the DM world,   and therefore we must consider here such issue.

\subsection*{Slides 63-65 }

DM is actively searched in a direct and
indirect way.  The distribution of dark matter in our galaxy, so as in
nearby ones,  plays an important role  in it.

Among the many candidates suggested for dark matter but as yet
undetected, one  among the most compelling possibility  is that the dark
matter is comprised of Weakly Interacting Massive Particles, or WIMPs,
a general class of particles, thermally produced in the hot early
universe, which drop out of equilibrium when they are not
relativistic. \citet{2005PhR...405..279B}  discuss  the  theoretical motivations and 
introduce a wide array of candidates for particle dark matter. 

 Its  density today is inversely proportional to their
annihilation rate, and in order to represent the `` CMB observed'' $\sim 25\%$ of the
critical density of the Universe, the particle's annihilation cross section should be typical
of electroweak-scale interactions, hinting at physics beyond the
Standard Model.  The  existence of such a particle  has plenty of theoretical
justification.  In fact, several extensions to the Standard Model lead
to WIMP candidates. One of them is Supersymmetry (SUSY), which extends
the Standard Model to include a new set of particles and interactions
that solves the gauge hierarchy problem, leads to a unification of the
coupling constants, and  it is required by string theory. The lightest
neutral SUSY particle, the neutralino, is thought to be stable and is
a natural dark matter candidate.

Direct detection of WIMPs requires sensing scattering between WIMPs in
the halo and atomic nuclei in a terrestrial detector.
Despite the
expected weak-scale coupling strength for such interactions,
terrestrial experiments are made feasible by the fact  that:   i) the local
DM mass-energy density is $10^5$ times higher than that of the universal
average. In detail, let us stress that  $\rho_{H}(R_\odot )$, the DM density at the Sun location,  has been recently   estimated it in an independent way  by \citet{2010A&A...523A..83S}:  $ \rho_{DM}(R_\odot) = (0.43 \pm 0.1 ) \ Gev \ cm ^{-3}  $  a value somewhat larger than that of   $0.3 \ GeV \ cm^{-3}$ usually assumed.  ii) coherence effects may amplify the interaction
rate (e.g. spin-independent nuclear elastic scattering is proportional
to $A^2$ the atomic weight of the target).

Even though direct detection may be possible, it is by no means
easy. The energy of the resulting nuclear recoils is likely of order
20 keV, where electromagnetic backgrounds dominate by many orders of
magnitude.  Nuclear collisions due to neutrons also pose a serious
obstacle, as they may exactly mimic those due to WIMPs. Hence, the
fundamental challenge of a direct search is to identify extremely rare
DM-induced nuclear recoil events from a vast array of spurious
signals. This can be achieved by combining low-radioactivity materials
and environments with robust background rejection of electron recoil
events. These detectors must also be situated in deep underground
laboratories to shield from cosmic-ray-induced backgrounds.

The dark matter community has developed a broad range of techniques to
address these challenges. Many different experiments are currently pursuing
one or more of the multiple possible detection channels. The most common
technique is to look for ionization and/or scintillation of the detector
medium (e.g. NaI, CsI, Ge, liquid/gaseous Xe or Ar) resulting from a
WIMP-nucleon collision. Other experiments look for phonons, bubble
nucleation, directionality effects.

About $50\%$ of all direct search experiments use noble liquids
(including XENON, LUX, XMASS, ZEPLIN, WARP, ArDM, and
DEEP/CLEAN). Noble liquids, especially Xe and Ar, are promising as a
detection medium as they are efficient scintillators, are chemically
stable, and have relatively high stopping power due to high mass and
density.

At present, the upper limit on the WIMP-nucleon cross section
approaches $10^{-44}$cm$^2$, which is well into the region of
parameter space where SUSY particles could account for the dark
matter. The next 2-3 orders of magnitude represent a particularly rich
region of electroweak-scale physics. The next generation of  direct search
experiments are poised to verify or reject a large portion of the
current theoretical explanations of WIMP-like DM.
 
\subsection*{Slides 66-68 }

DM particles makes  the galactic
halo where they would continuously annihilate into quark or gauge
boson pairs  leading eventually to rare antimatter particles
and high-energy photons and neutrinos.The indirect serches of DM are based on astrophysical observations
of the products of DM self annihilation or decay.
Given the known long lifetime of the DM, the signal for decay
products is suppressed for heavy candidates (due to the combination of
low number densities and long lifetime) leaving in most of the cases
only the
self-annihilation as most sensitive possible source of a signal.
Antimatter particles such as
antiprotons and positrons are interesting because they are quite rare in Nature. Any
(additional) production  through DM annihilation will induce  an excess
in the energy distributions of the flux observed at the earth. 
However,  let us stress that  conventional
astrophysical mechanisms in the galaxy also produce antiprotons and
positrons.  
 
The indirect searches of DM are based on astrophysical observations
of the products of DM self annihilation or decay.
Given the known long lifetime of the DM, the signal for decay
products is suppressed for heavy candidates (due to the combination of
low number densities and long lifetime) leaving in most of the cases
only the
self-annihilation as most sensitive possible source of a signal.

In the case of searches via gamma ray observation,
the expected flux in a detector on Earth is given by:
\begin{equation}
\frac{d \phi_{\gamma}}{d E_{\gamma}} (E_{\gamma} , \Delta \psi)
=\frac{\langle \sigma v \rangle_{ann}}{4 \pi m2_{\chi}} \sum_f B_{f}
\frac{dN^f_ {\gamma}}{d E_ {\gamma}}
\times
\frac{1}{2} \int_{\Delta  \psi}\frac{d\Omega}{\Delta \psi}
\int_{l.o.s.} dl (\psi) \rho^2 (r),
\end{equation}
where $E_{\gamma}$ is the photon energy, $m_{\chi}$ is the DM particle
mass,
$\Delta \psi$ is the detector opening angle,
$\langle \sigma v \rangle _{ann}$ is the mean annihilation cross
section times the relative velocity (of order 10$^{-26}$ cm$^3$
s$^{-1}$ for cold WIMP relics from abundances constrains),
$B_f$ indicates the branching fraction in a given channel $f$,
$\frac{dN^f_{\gamma}}{dE_{\gamma}}$ is the photon spectrum for a given
annihilation channel which depends on the DM model and can have both
continuum and discrete lines contributions, $\rho$ is the DM density
and the integrals are along the line of sight and over the detector
angle.

The quadratic dependence on $\rho$ suggest that the preferred targets
for indirect searches are the places with higher DM concentrations,
like the centre of galaxies, satellites or galaxy clusters.
It must be noticed however that the galactic centres are very often
sources of strong activities due for example of the presence of black
holes or other compact objects enhancing the overall
background.
Moreover the large uncertainty on the DM density profile directly
reflects on the flux predictions making the searches extremely
difficult (although possible enhancements due to local DM
over-densities are possible).

DM particles can also generate high-energy photons.  The Fermi satellite for example maybe able to indirectly detect DM 
particles looking at high energy photons.As example let us consider  DM species  at 40 GeV WIMP from a
supersymmetric theory. That particle is supposed to annihilate into
bottom- antibottom quark pairs.
 In this scenario, we show in the slide the flux  map generated after five years of  satellite operation.
The map  features the DM signal to noise ratio. The DM signal is the number of gamma
rays produced towards each pixel by the DM distribution. The smooth DM distribution and  
the heaviest and closest DM substructures (bright spots on the maps),  as expected in $\Lambda CDM $ are considered.

The noise includes the galactic diffuse emission mostly arising from the interactions of cosmic rays
protons and helium nuclei on the HI and H$_2$ gas components of the
Milky Way. The two maps refer to different set of 
simulations: Aquarius (left panel) and Via Lactea II (right
panel). 
 It is then  obvious how the DM density distribution in galaxies  $\rho_{H}$ plays a role, in particular, according whether in the innermost $R_{vir}/20$ region  it is cuspy  $\rho_{H}(r) \propto r^{-1}$ or constant. The predicted flux can differ by several orders of magnitude.  Furthermore, one should also take into account the flux  enhancements due to local DM over-densities i.e. sub-halo, but this is a very open question.

 Another possibility is to pursue indirect search by looking at charged
particles such as positrons or antiprotons, in this case however,  the
galactic magnetic field is such that the direction of arrival of the
particle does not reflect the production point and the only observable
is an excess of antimatter with respect to the expected background due
to ordinary cosmic rays (which also suffer from big uncertainties).

Among the most important facilities for indirect signals we
have: XMM Newton and Chandra, Integral, Compton Gamma Ray Observatory,
AGILE, Fermi-LAT Observatory, CANGAROO, HESS, MAGIC, VERITAS, AMANDA,
ICECUBE, ANTARES, PAMELA and AMS.

Up today several claims of indirect DM detection has been made,
sometimes in conflict with each other and all of them competing with
other astrophysical explanations, 
The most significant are:
the positron excess measured by HEAT and PAMELA, the 511 keV line excess measured
by INTEGRAL, the EGRET Diffuse Galactic Spectrum and the the so called
WMAP Haze (an excess of microwave emission around the centre of the
Milky Way) see  \citet{2010MmSAI..81..123M}  for the current status
on these searches.
.

\section{ What we know?}
\subsection*{Slides 69}

The distribution of DM halos around galaxies shows a striking and complex phenomenology. We believe that it is giving invaluable information on the Nature itself of dark matter  and on the galaxy formation processes.

Semi-analytical models as Baryon +DM simulations should reproduce the empirical scenario:
a shallow DM inner distribution, strong relationships between the  halo mass and 1) central halo density, 2) baryonic mass, 3) half-mass baryonic radius and 4) baryonic central surface density, a constant central halo surface density.
In any case,  the mass discrepancy  in  the World-Universe of galaxies is a clear function  of radius, total baryonic mass and  Hubble Type
$$
 \frac {M_{grav}} {M_{b}} \sim  \frac {1 + \gamma_1(M_b, T) r^3} {1 + \gamma_2 (M_b, T) r^2}
$$
that very unlikely may arise as effect of a change of the  gravitation law.

\newpage

\newpage

\bibliographystyle{AA}
\small
\baselineskip=0ex 
\bibliography{dmaw2010}

\begin{thebibliography}{83}
\expandafter\ifx\csname natexlab\endcsname\relax\def\natexlab#1{#1}\fi

\bibitem[{{Aaronson}(1983)}]{1983ApJ...266L..11A}
{Aaronson}, M. 1983, \apjl, 266, L11

\bibitem[{{Bell} \& {de Jong}(2001)}]{2001ApJ...550..212B}
{Bell}, E.~F. \& {de Jong}, R.~S. 2001, \apj, 550, 212

\bibitem[{{Belokurov} {et~al.}(2007){Belokurov}, {Zucker}, {Evans}, {Kleyna},
  {Koposov}, {Hodgkin}, {Irwin}, {Gilmore}, {Wilkinson}, {Fellhauer},
  {Bramich}, {Hewett}, {Vidrih}, {De Jong}, {Smith}, {Rix}, {Bell}, {Wyse},
  {Newberg}, {Mayeur}, {Yanny}, {Rockosi}, {Gnedin}, {Schneider}, {Beers},
  {Barentine}, {Brewington}, {Brinkmann}, {Harvanek}, {Kleinman}, {Krzesinski},
  {Long}, {Nitta}, \& {Snedden}}]{2007ApJ...654..897B}
{Belokurov}, V., {Zucker}, D.~B., {Evans}, N.~W., {et~al.} 2007, \apj, 654, 897

\bibitem[{{Bertone} {et~al.}(2005){Bertone}, {Hooper}, \&
  {Silk}}]{2005PhR...405..279B}
{Bertone}, G., {Hooper}, D., \& {Silk}, J. 2005, \physrep, 405, 279

\bibitem[{{Bosma}(1978)}]{1978PhDT.......195B}
{Bosma}, A. 1978, PhD thesis, PhD Thesis, Groningen Univ., (1978)

\bibitem[{{Bosma}(1981{\natexlab{a}})}]{1981AJ.....86.1791B}
{Bosma}, A. 1981{\natexlab{a}}, \aj, 86, 1791

\bibitem[{{Bosma}(1981{\natexlab{b}})}]{1981AJ.....86.1825B}
{Bosma}, A. 1981{\natexlab{b}}, \aj, 86, 1825

\bibitem[{{Bosma} \& {van der Kruit}(1979)}]{1979A&A....79..281B}
{Bosma}, A. \& {van der Kruit}, P.~C. 1979, \aap, 79, 281

\bibitem[{{Burkert}(1995)}]{1995ApJ...447L..25B}
{Burkert}, A. 1995, \apjl, 447, L25+

\bibitem[{{Cappellari}(2008)}]{2008MNRAS.390...71C}
{Cappellari}, M. 2008, \mnras, 390, 71

\bibitem[{{Cappellari} {et~al.}(2006){Cappellari}, {Bacon}, {Bureau}, {Damen},
  {Davies}, {de Zeeuw}, {Emsellem}, {Falc{\'o}n-Barroso}, {Krajnovi{\'c}},
  {Kuntschner}, {McDermid}, {Peletier}, {Sarzi}, {van den Bosch}, \& {van de
  Ven}}]{2006MNRAS.366.1126C}
{Cappellari}, M., {Bacon}, R., {Bureau}, M., {et~al.} 2006, \mnras, 366, 1126

\bibitem[{{Catinella} {et~al.}(2006){Catinella}, {Giovanelli}, \&
  {Haynes}}]{2006ApJ...640..751C}
{Catinella}, B., {Giovanelli}, R., \& {Haynes}, M.~P. 2006, \apj, 640, 751

\bibitem[{{Coccato} {et~al.}(2008){Coccato}, {Gerhard}, {Arnaboldi}, {Das},
  {Douglas}, {Kuijken}, {Merrifield}, {Napolitano}, {Noordermeer},
  {Romanowsky}, {Capaccioli}, {Cortesi}, {De Lorenzi}, \&
  {Freeman}}]{2008AN....329..912C}
{Coccato}, L., {Gerhard}, O., {Arnaboldi}, M., {et~al.} 2008, Astronomische
  Nachrichten, 329, 912

\bibitem[{{Coccato} {et~al.}(2009){Coccato}, {Gerhard}, {Arnaboldi}, {Das},
  {Douglas}, {Kuijken}, {Merrifield}, {Napolitano}, {Noordermeer},
  {Romanowsky}, {Capaccioli}, {Cortesi}, {de Lorenzi}, \&
  {Freeman}}]{2009MNRAS.394.1249C}
{Coccato}, L., {Gerhard}, O., {Arnaboldi}, M., {et~al.} 2009, \mnras, 394, 1249

\bibitem[{{Conroy} {et~al.}(2006){Conroy}, {Wechsler}, \&
  {Kravtsov}}]{2006ApJ...647..201C}
{Conroy}, C., {Wechsler}, R.~H., \& {Kravtsov}, A.~V. 2006, \apj, 647, 201

\bibitem[{{C{\^o}t{\'e}} {et~al.}(1999){C{\^o}t{\'e}}, {Mateo}, {Olszewski}, \&
  {Cook}}]{1999ApJ...526..147C}
{C{\^o}t{\'e}}, P., {Mateo}, M., {Olszewski}, E.~W., \& {Cook}, K.~H. 1999,
  \apj, 526, 147

\bibitem[{{de Blok} {et~al.}(2008){de Blok}, {Walter}, {Brinks},
  {Trachternach}, {Oh}, \& {Kennicutt}}]{2008AJ....136.2648D}
{de Blok}, W.~J.~G., {Walter}, F., {Brinks}, E., {et~al.} 2008, \aj, 136, 2648

\bibitem[{{Donato} {et~al.}(2009){Donato}, {Gentile}, {Salucci}, {Frigerio
  Martins}, {Wilkinson}, {Gilmore}, {Grebel}, {Koch}, \&
  {Wyse}}]{2009MNRAS.397.1169D}
{Donato}, F., {Gentile}, G., {Salucci}, P., {et~al.} 2009, \mnras, 397, 1169

\bibitem[{{Douglas} {et~al.}(2002){Douglas}, {Arnaboldi}, {Freeman}, {Kuijken},
  {Merrifield}, {Romanowsky}, {Taylor}, {Capaccioli}, {Axelrod}, {Gilmozzi},
  {Hart}, {Bloxham}, \& {Jones}}]{2002PASP..114.1234D}
{Douglas}, N.~G., {Arnaboldi}, M., {Freeman}, K.~C., {et~al.} 2002, \pasp, 114,
  1234

\bibitem[{{Ettori} \& {Fabian}(2006)}]{2006MNRAS.369L..42E}
{Ettori}, S. \& {Fabian}, A.~C. 2006, \mnras, 369, L42

\bibitem[{{Fabricant} {et~al.}(1984){Fabricant}, {Rybicki}, \&
  {Gorenstein}}]{1984ApJ...286..186F}
{Fabricant}, D., {Rybicki}, G., \& {Gorenstein}, P. 1984, \apj, 286, 186

\bibitem[{{Freeman}(1970)}]{1970ApJ...160..811F}
{Freeman}, K.~C. 1970, \apj, 160, 811

\bibitem[{{Gargiulo} {et~al.}(2009){Gargiulo}, {Haines}, {Merluzzi}, {Smith},
  {Barbera}, {Busarello}, {Lucey}, {Mercurio}, \&
  {Capaccioli}}]{2009MNRAS.397...75G}
{Gargiulo}, A., {Haines}, C.~P., {Merluzzi}, P., {et~al.} 2009, \mnras, 397, 75

\bibitem[{{Gentile} {et~al.}(2005){Gentile}, {Burkert}, {Salucci}, {Klein}, \&
  {Walter}}]{2005ApJ...634L.145G}
{Gentile}, G., {Burkert}, A., {Salucci}, P., {Klein}, U., \& {Walter}, F. 2005,
  \apjl, 634, L145

\bibitem[{{Gentile} {et~al.}(2004){Gentile}, {Salucci}, {Klein}, {Vergani}, \&
  {Kalberla}}]{2004MNRAS.351..903G}
{Gentile}, G., {Salucci}, P., {Klein}, U., {Vergani}, D., \& {Kalberla}, P.
  2004, \mnras, 351, 903

\bibitem[{{Gilmore} {et~al.}(2007{\natexlab{a}}){Gilmore}, {Wilkinson},
  {Kleyna}, {Koch}, {Evans}, {Wyse}, \& {Grebel}}]{2007NuPhS.173...15G}
{Gilmore}, G., {Wilkinson}, M., {Kleyna}, J., {et~al.} 2007{\natexlab{a}},
  Nuclear Physics B Proceedings Supplements, 173, 15

\bibitem[{{Gilmore} {et~al.}(2007{\natexlab{b}}){Gilmore}, {Wilkinson}, {Wyse},
  {Kleyna}, {Koch}, {Evans}, \& {Grebel}}]{2007ApJ...663..948G}
{Gilmore}, G., {Wilkinson}, M.~I., {Wyse}, R.~F.~G., {et~al.}
  2007{\natexlab{b}}, \apj, 663, 948

\bibitem[{{Gnedin} {et~al.}(2004){Gnedin}, {Kravtsov}, {Klypin}, \&
  {Nagai}}]{2004ApJ...616...16G}
{Gnedin}, O.~Y., {Kravtsov}, A.~V., {Klypin}, A.~A., \& {Nagai}, D. 2004, \apj,
  616, 16

\bibitem[{{Governato} {et~al.}(2010){Governato}, {Brook}, {Mayer}, {Brooks},
  {Rhee}, {Wadsley}, {Jonsson}, {Willman}, {Stinson}, {Quinn}, \&
  {Madau}}]{2010Natur.463..203G}
{Governato}, F., {Brook}, C., {Mayer}, L., {et~al.} 2010, \nat, 463, 203

\bibitem[{{Grillo} {et~al.}(2009){Grillo}, {Gobat}, {Lombardi}, \&
  {Rosati}}]{2009A&A...501..461G}
{Grillo}, C., {Gobat}, R., {Lombardi}, M., \& {Rosati}, P. 2009, \aap, 501, 461

\bibitem[{{Hyde} \& {Bernardi}(2009)}]{2009MNRAS.396.1171H}
{Hyde}, J.~B. \& {Bernardi}, M. 2009, \mnras, 396, 1171

\bibitem[{{Jorgensen} {et~al.}(1996){Jorgensen}, {Franx}, \&
  {Kjaergaard}}]{1996MNRAS.280..167J}
{Jorgensen}, I., {Franx}, M., \& {Kjaergaard}, P. 1996, \mnras, 280, 167

\bibitem[{{Klypin} {et~al.}(2010){Klypin}, {Trujillo-Gomez}, \&
  {Primack}}]{2010arXiv1002.3660K}
{Klypin}, A., {Trujillo-Gomez}, S., \& {Primack}, J. 2010, ArXiv e-prints

\bibitem[{{Koopmans} {et~al.}(2006){Koopmans}, {Treu}, {Bolton}, {Burles}, \&
  {Moustakas}}]{2006ApJ...649..599K}
{Koopmans}, L.~V.~E., {Treu}, T., {Bolton}, A.~S., {Burles}, S., \&
  {Moustakas}, L.~A. 2006, \apj, 649, 599

\bibitem[{{Kormendy} \& {Freeman}(2004)}]{2004IAUS..220..377K}
{Kormendy}, J. \& {Freeman}, K.~C. 2004, in IAU Symposium, Vol. 220, Dark
  Matter in Galaxies, ed. {S.~Ryder, D.~Pisano, M.~Walker, \& K.~Freeman},
  377--+

\bibitem[{{Krajnovi{\'c}} {et~al.}(2005){Krajnovi{\'c}}, {Cappellari},
  {Emsellem}, {McDermid}, \& {de Zeeuw}}]{2005MNRAS.357.1113K}
{Krajnovi{\'c}}, D., {Cappellari}, M., {Emsellem}, E., {McDermid}, R.~M., \&
  {de Zeeuw}, P.~T. 2005, \mnras, 357, 1113

\bibitem[{{Kuzio de Naray} {et~al.}(2008){Kuzio de Naray}, {McGaugh}, \& {de
  Blok}}]{2008ApJ...676..920K}
{Kuzio de Naray}, R., {McGaugh}, S.~S., \& {de Blok}, W.~J.~G. 2008, \apj, 676,
  920

\bibitem[{{Lapi} \& {Cavaliere}(2009)}]{2009ApJ...692..174L}
{Lapi}, A. \& {Cavaliere}, A. 2009, \apj, 692, 174

\bibitem[{{Mamon} \& {{\L}okas}(2005)}]{2005MNRAS.362...95M}
{Mamon}, G.~A. \& {{\L}okas}, E.~L. 2005, \mnras, 362, 95

\bibitem[{{Mandelbaum} {et~al.}(2006){Mandelbaum}, {Seljak}, {Kauffmann},
  {Hirata}, \& {Brinkmann}}]{2006MNRAS.368..715M}
{Mandelbaum}, R., {Seljak}, U., {Kauffmann}, G., {Hirata}, C.~M., \&
  {Brinkmann}, J. 2006, \mnras, 368, 715

\bibitem[{{Mandelbaum} {et~al.}(2009){Mandelbaum}, {van de Ven}, \&
  {Keeton}}]{2009MNRAS.398..635M}
{Mandelbaum}, R., {van de Ven}, G., \& {Keeton}, C.~R. 2009, \mnras, 398, 635

\bibitem[{{Maraston}(1998)}]{1998MNRAS.300..872M}
{Maraston}, C. 1998, \mnras, 300, 872

\bibitem[{{Maraston}(2005)}]{2005MNRAS.362..799M}
{Maraston}, C. 2005, \mnras, 362, 799

\bibitem[{{Maraston} {et~al.}(2006){Maraston}, {Daddi}, {Renzini}, {Cimatti},
  {Dickinson}, {Papovich}, {Pasquali}, \& {Pirzkal}}]{2006ApJ...652...85M}
{Maraston}, C., {Daddi}, E., {Renzini}, A., {et~al.} 2006, \apj, 652, 85

\bibitem[{{Marinoni} \& {Hudson}(2002)}]{2002ApJ...569..101M}
{Marinoni}, C. \& {Hudson}, M.~J. 2002, \apj, 569, 101

\bibitem[{{Mateo}(1997)}]{1997ASPC..116..259M}
{Mateo}, M. 1997, in Astronomical Society of the Pacific Conference Series,
  Vol. 116, The Nature of Elliptical Galaxies; 2nd Stromlo Symposium, ed.
  {M.~Arnaboldi, G.~S.~Da Costa, \& P.~Saha}, 259--+

\bibitem[{{Mateo}(1998)}]{1998ARA&A..36..435M}
{Mateo}, M.~L. 1998, \araa, 36, 435

\bibitem[{{M{\'e}ndez} {et~al.}(2009){M{\'e}ndez}, {Teodorescu}, {Kudritzki},
  \& {Burkert}}]{2009ApJ...691..228M}
{M{\'e}ndez}, R.~H., {Teodorescu}, A.~M., {Kudritzki}, R., \& {Burkert}, A.
  2009, \apj, 691, 228

\bibitem[{{Mo} {et~al.}(1998){Mo}, {Mao}, \& {White}}]{1998MNRAS.295..319M}
{Mo}, H.~J., {Mao}, S., \& {White}, S.~D.~M. 1998, \mnras, 295, 319

\bibitem[{{Morselli}(2010)}]{2010MmSAI..81..123M}
{Morselli}, A. 2010, \memsai, 81, 123

\bibitem[{{Moster} {et~al.}(2010){Moster}, {Somerville}, {Maulbetsch}, {van den
  Bosch}, {Macci{\`o}}, {Naab}, \& {Oser}}]{2010ApJ...710..903M}
{Moster}, B.~P., {Somerville}, R.~S., {Maulbetsch}, C., {et~al.} 2010, \apj,
  710, 903

\bibitem[{{Nagino} \& {Matsushita}(2009)}]{2009A&A...501..157N}
{Nagino}, R. \& {Matsushita}, K. 2009, \aap, 501, 157

\bibitem[{{Napolitano} {et~al.}(2001){Napolitano}, {Arnaboldi}, {Freeman}, \&
  {Capaccioli}}]{2001A&A...377..784N}
{Napolitano}, N.~R., {Arnaboldi}, M., {Freeman}, K.~C., \& {Capaccioli}, M.
  2001, \aap, 377, 784

\bibitem[{{Napolitano} {et~al.}(2010){Napolitano}, {Romanowsky}, {Capaccioli},
  {Douglas}, {Arnaboldi}, {Coccato}, {Gerhard}, {Kuijken}, {Merrifield},
  {Bamford}, {Cortesi}, {Das}, \& {Freeman}}]{2010MNRAS.tmp.1835N}
{Napolitano}, N.~R., {Romanowsky}, A.~J., {Capaccioli}, M., {et~al.} 2010,
  \mnras, 1835

\bibitem[{{Navarro} {et~al.}(1997){Navarro}, {Frenk}, \&
  {White}}]{1997ApJ...490..493N}
{Navarro}, J.~F., {Frenk}, C.~S., \& {White}, S.~D.~M. 1997, \apj, 490, 493

\bibitem[{{Oh} {et~al.}(2008){Oh}, {de Blok}, {Walter}, {Brinks}, \&
  {Kennicutt}}]{2008AJ....136.2761O}
{Oh}, S., {de Blok}, W.~J.~G., {Walter}, F., {Brinks}, E., \& {Kennicutt},
  R.~C. 2008, \aj, 136, 2761

\bibitem[{{Persic} {et~al.}(1996){Persic}, {Salucci}, \&
  {Stel}}]{1996MNRAS.281...27P}
{Persic}, M., {Salucci}, P., \& {Stel}, F. 1996, \mnras, 281, 27

\bibitem[{{Plummer}(1915)}]{1915MNRAS..76..107P}
{Plummer}, H.~C. 1915, \mnras, 76, 107

\bibitem[{{Rubin} {et~al.}(1980){Rubin}, {Peterson}, \&
  {Ford}}]{1980ApJ...239...50R}
{Rubin}, V.~C., {Peterson}, C.~J., \& {Ford}, Jr., W.~K. 1980, \apj, 239, 50

\bibitem[{{Salucci} {et~al.}(2007){Salucci}, {Lapi}, {Tonini}, {Gentile},
  {Yegorova}, \& {Klein}}]{2007MNRAS.378...41S}
{Salucci}, P., {Lapi}, A., {Tonini}, C., {et~al.} 2007, \mnras, 378, 41

\bibitem[{{Salucci} {et~al.}(2010){Salucci}, {Nesti}, {Gentile}, \& {Frigerio
  Martins}}]{2010A&A...523A..83S}
{Salucci}, P., {Nesti}, F., {Gentile}, G., \& {Frigerio Martins}, C. 2010,
  \aap, 523, A83+

\bibitem[{{Schneider}(1996)}]{1996MNRAS.283..837S}
{Schneider}, P. 1996, \mnras, 283, 837

\bibitem[{{Schneider, P., Kochanek, C.~S., \& Wambsganss,
  J.}(2006)}]{2006glsw.book..269S}
{Schneider, P., Kochanek, C.~S., \& Wambsganss, J.}, ed. 2006, {Weak
  Gravitational Lensing} (Springer-Verlag), 269--+

\bibitem[{{Shankar} \& {Bernardi}(2009)}]{2009MNRAS.396L..76S}
{Shankar}, F. \& {Bernardi}, M. 2009, \mnras, 396, L76

\bibitem[{{Shankar} {et~al.}(2006){Shankar}, {Lapi}, {Salucci}, {De Zotti}, \&
  {Danese}}]{2006ApJ...643...14S}
{Shankar}, F., {Lapi}, A., {Salucci}, P., {De Zotti}, G., \& {Danese}, L. 2006,
  \apj, 643, 14

\bibitem[{{Spano} {et~al.}(2008){Spano}, {Marcelin}, {Amram}, {Carignan},
  {Epinat}, \& {Hernandez}}]{2008MNRAS.383..297S}
{Spano}, M., {Marcelin}, M., {Amram}, P., {et~al.} 2008, \mnras, 383, 297

\bibitem[{{Strigari} {et~al.}(2008){Strigari}, {Bullock}, {Kaplinghat},
  {Simon}, {Geha}, {Willman}, \& {Walker}}]{2008Natur.454.1096S}
{Strigari}, L.~E., {Bullock}, J.~S., {Kaplinghat}, M., {et~al.} 2008, \nat,
  454, 1096

\bibitem[{{Tinsley}(1975)}]{1975MmSAI..46....3T}
{Tinsley}, B.~M. 1975, \memsai, 46, 3

\bibitem[{{Tiret} {et~al.}(2010){Tiret}, {Salucci}, {Bernardi}, {Maraston}, \&
  {Pforr}}]{2010MNRAS.tmp.1737T}
{Tiret}, O., {Salucci}, P., {Bernardi}, M., {Maraston}, C., \& {Pforr}, J.
  2010, \mnras, 1737

\bibitem[{{Trachternach} {et~al.}(2008){Trachternach}, {de Blok}, {Walter},
  {Brinks}, \& {Kennicutt}}]{2008AJ....136.2720T}
{Trachternach}, C., {de Blok}, W.~J.~G., {Walter}, F., {Brinks}, E., \&
  {Kennicutt}, R.~C. 2008, \aj, 136, 2720

\bibitem[{{Tully} \& {Fisher}(1977)}]{1977A&A....54..661T}
{Tully}, R.~B. \& {Fisher}, J.~R. 1977, \aap, 54, 661

\bibitem[{{Vale} \& {Ostriker}(2004)}]{2004MNRAS.353..189V}
{Vale}, A. \& {Ostriker}, J.~P. 2004, \mnras, 353, 189

\bibitem[{{Walker} {et~al.}(2007){Walker}, {Mateo}, {Olszewski}, {Gnedin},
  {Wang}, {Sen}, \& {Woodroofe}}]{2007ApJ...667L..53W}
{Walker}, M.~G., {Mateo}, M., {Olszewski}, E.~W., {et~al.} 2007, \apjl, 667,
  L53

\bibitem[{{Walker} {et~al.}(2009){Walker}, {Mateo}, {Olszewski},
  {Pe{\~n}arrubia}, {Wyn Evans}, \& {Gilmore}}]{2009ApJ...704.1274W}
{Walker}, M.~G., {Mateo}, M., {Olszewski}, E.~W., {et~al.} 2009, \apj, 704,
  1274

\bibitem[{{Walker} {et~al.}(2010{\natexlab{a}}){Walker}, {Mateo}, {Olszewski},
  {Pe{\~n}arrubia}, {Wyn Evans}, \& {Gilmore}}]{2010ApJ...710..886W}
{Walker}, M.~G., {Mateo}, M., {Olszewski}, E.~W., {et~al.} 2010{\natexlab{a}},
  \apj, 710, 886

\bibitem[{{Walker} {et~al.}(2010{\natexlab{b}}){Walker}, {McGaugh}, {Mateo},
  {Olszewski}, \& {Kuzio de Naray}}]{2010ApJ...717L..87W}
{Walker}, M.~G., {McGaugh}, S.~S., {Mateo}, M., {Olszewski}, E.~W., \& {Kuzio
  de Naray}, R. 2010{\natexlab{b}}, \apjl, 717, L87

\bibitem[{{Wechsler} {et~al.}(2002){Wechsler}, {Bullock}, {Primack},
  {Kravtsov}, \& {Dekel}}]{2002ApJ...568...52W}
{Wechsler}, R.~H., {Bullock}, J.~S., {Primack}, J.~R., {Kravtsov}, A.~V., \&
  {Dekel}, A. 2002, \apj, 568, 52

\bibitem[{{Wechsler} {et~al.}(2006){Wechsler}, {Zentner}, {Bullock},
  {Kravtsov}, \& {Allgood}}]{2006ApJ...652...71W}
{Wechsler}, R.~H., {Zentner}, A.~R., {Bullock}, J.~S., {Kravtsov}, A.~V., \&
  {Allgood}, B. 2006, \apj, 652, 71

\bibitem[{{Wilkinson} {et~al.}(2006){Wilkinson}, {Kleyna}, {Wyn Evans},
  {Gilmore}, {Read}, {Koch}, {Grebel}, \& {Irwin}}]{2006EAS....20..105W}
{Wilkinson}, M.~I., {Kleyna}, J.~T., {Wyn Evans}, N., {et~al.} 2006, in EAS
  Publications Series, Vol.~20, EAS Publications Series, ed. {G.~A.~Mamon,
  F.~Combes, C.~Deffayet, \& B.~Fort}, 105--112

\bibitem[{{Willman} {et~al.}(2005){Willman}, {Blanton}, {West}, {Dalcanton},
  {Hogg}, {Schneider}, {Wherry}, {Yanny}, \& {Brinkmann}}]{2005AJ....129.2692W}
{Willman}, B., {Blanton}, M.~R., {West}, A.~A., {et~al.} 2005, \aj, 129, 2692

\bibitem[{{Wolf} {et~al.}(2010){Wolf}, {Martinez}, {Bullock}, {Kaplinghat},
  {Geha}, {Mu{\~n}oz}, {Simon}, \& {Avedo}}]{2010MNRAS.406.1220W}
{Wolf}, J., {Martinez}, G.~D., {Bullock}, J.~S., {et~al.} 2010, \mnras, 406,
  1220

\bibitem[{{Wong} \& {Blitz}(2002)}]{2002ApJ...569..157W}
{Wong}, T. \& {Blitz}, L. 2002, \apj, 569, 157

\bibitem[{{Yegorova} \& {Salucci}(2007)}]{2007MNRAS.377..507Y}
{Yegorova}, I.~A. \& {Salucci}, P. 2007, \mnras, 377, 507

\end{thebibliography}
\newpage

\begin{figure*}
\begin{center}
\psfig{file=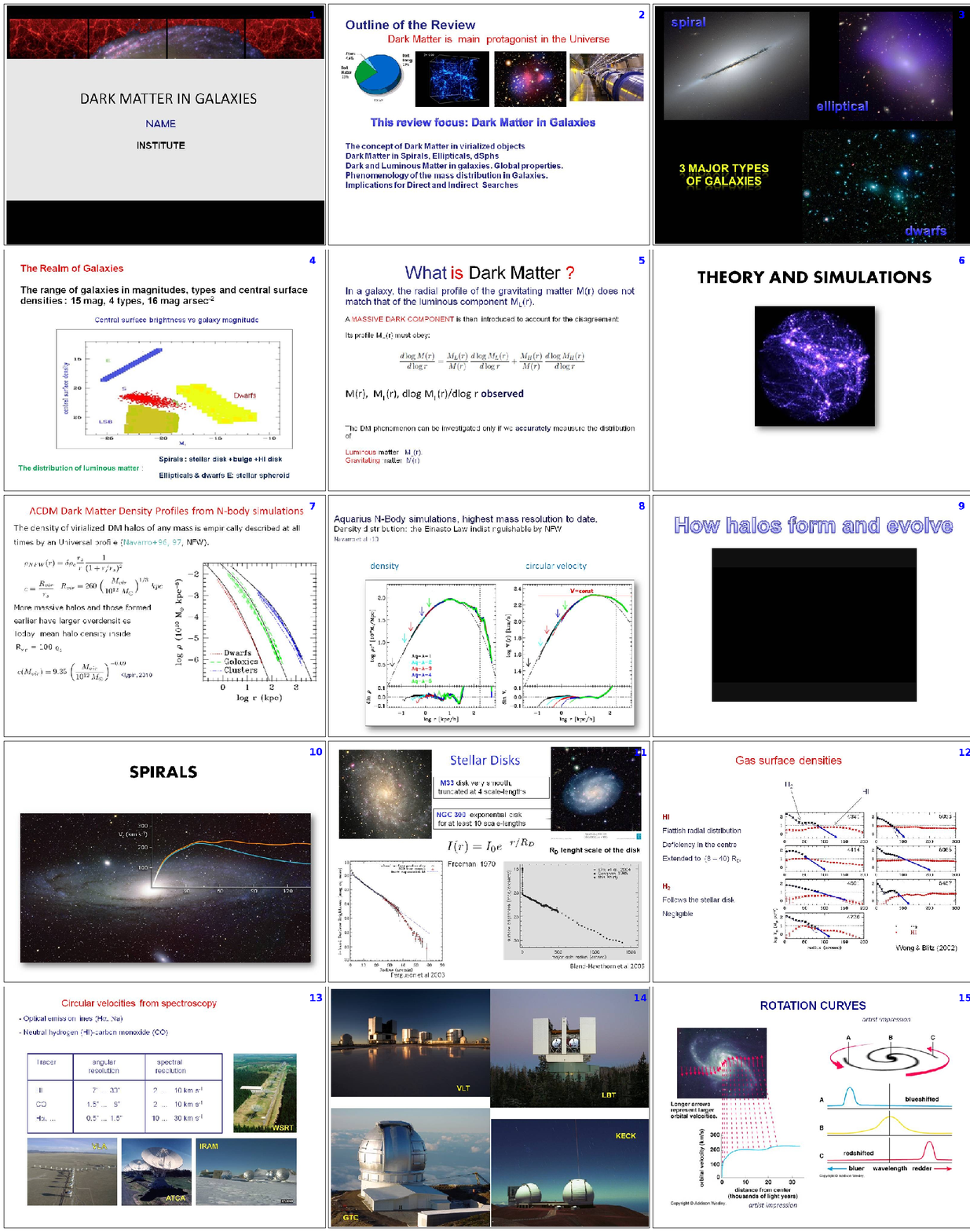 ,width=1.\textwidth}
\end{center}
\caption{Zoomed out of the first 15 slides of the Presentation Review, shown here to give a glimpse of it. These slides can be downloaded at \url{http://www.sissa.it.ap/dmg/dmaw_presentation.html}}

\end {figure*}
\newpage

\appendix
\section{APPENDIX  DMAW 2010 Seminar Speakers }

\scriptsize
\begin{center}
\begin{longtable}{|l |l| l| l|}

\hline
Speaker                 & Institute                           & City       & Nation    \\\hline                                                                                                        
S. AL-Jaber             & An-Najah National University        & Nablus     & Palestine    \\\hline                                                                                                        
J. Aleksic              & Institut de Fisica d'Altes Energies & Bellaterra & Spain\\\hline
G. Alves Silva          & Universidade Estadual de Roraima    & Boa Vista  & Brasil\\\hline
J. Bailin               & University of Michigan              & Ann Arbor  & USA\\\hline
M. Balcells             & Isaac Newton Group of Telescopes    & La Palma   & Spain\\\hline
M. Baldi                & Excellence Cluster Universe         & Garching   & Germany\\\hline
G. Barbiellini          & INFN Trieste                        & Trieste    & Italy\\\hline
N. Bartolo              & Physics Department ``Galileo Galilei'', University of Padova $\&$ & Padova & Italy\\ 
                        & INFN Padova                         & Padova     & Italy\\\hline
M. Bastero Gil          & University of Granada               & Granada    & Spain\\\hline
G. K. Beeharry          & Mauritius Radio Telescope, University of Mauritius & Reduit & Mauritius\\\hline
Bernal, Degollado, Hidalgo, & Universidad Nacional Autonoma de Mexico & Mexico City & Mexico\\
Barranco $\&$ Mastache   &                                     &            & \\\hline
M. Bernardi             & University of Pennsylvania          & Philadelphia& USA\\\hline
O. Bertolami            & University of Porto                 & Porto      & Portugal\\\hline
N. Bilic                & Rudjer Boskovic Institute           & Zagreb     & Croatia\\
                        & University of Zagreb                & Zagreb     & Croatia\\\hline
S. Bird                 & University of Turku                 & Turku      & Finland\\\hline
A. Biviano              & Osservatorio Astronomico di Trieste & Trieste    & Italy\\\hline
C. B\"{o}ehmer          & Institute of Origins, University College London  & London & UK\\\hline
A. Bosma                & Beijing Normal University $\&$           & Beijing    & China\\
                        & Laboratoire d'Astrophysique de Marseille& Marseille& France\\\hline
F. Bournaud             & CEA Saclay                          & Saclay     & France\\\hline
H. Bray $\&$ A. Badin   & Duke University                     & Durham     & USA\\\hline
A. Bressan              & Osservatorio Astronomico di Padova $\&$ Dept. Astronomy & Padua & Italy\\\hline
I. Brown                & University of Oslo                  & Oslo       & Norway\\\hline
D. Cabral Rodrigues     & Universidade Federal do Esp\'{i}rito Santo & Vit\'{o}ria & Brazil\\\hline
S. Capozziello          & Universit\`{a} di Napoli "Federico II"& Naples   & Italy\\\hline
L. Caramete $\&$ P. Stefanescu & Institute for Space Sciences & Bucharest-Magurele & Romania\\\hline 
B. Cervantes-Sodi       & Korea Astronomy $\&$ Space Science Institute & Daejeon & Korea\\\hline
R. Chavez               & INAOE Instituto Nacional de Astrofisica Optica y Electronica & Puebla & Mexico\\\hline
L. Chemin               & Laboratoire d'Astrophysique de Bordeaux & Floirac & France\\\hline
P. Chimenti             & UFABC Universidade Federal do ABC   & Santo Andr\'{e} & Brazil\\\hline
S. Colafrancesco        & Osservatorio Astronomico di Roma $\&$   & Roma       & Italy\\
                        & Universit\`{a} di Roma La Sapienza  & Roma       & Italy\\\hline
L.P. Colatto            & CEFET Petr\'{o}polis                & Petr\'{o}polis & Brazil\\\hline                   
C. Collins              & Liverpool John Moores University    & Merseyside & UK\\\hline
M. Colpi                & University of Milano Bicocca, Physics Dept. G. Occhialini & Milan & Italy\\\hline
G. Dastegir Al-Quaderi  & University of Dhaka                 & Dhaka      & Bangladesh\\\hline
R. Dave                 & Raman Research Institute            & Bangalore  & India\\\hline
R. de Grijs             & Kavli Institute for Astronomy $\&$ Astrophysics, Peking University& Beijing & China\\\hline
F. De Paolis            & University of Salento               & Lecce      & Italy\\\hline
A. Del Popolo           & Catania University                  & Catania    & Italy\\\hline
S. di Serego Alighieri  & Osservatorio Astrofisico di Arcetri & Florence   & Italy\\\hline
A. Erkurt               & University of the Basque Country    & Bizkaia    & Spain\\\hline
M. Fairbairn            & King's College                      & London     & UK\\\hline 
B. Famaey               & Observatoire Astronomique de Strasbourg & Strasbourg & France\\\hline
V. Faraoni              & Bishop's University                 & Sherbrooke & Canada\\\hline
R. Fisher               & UMass Dartmouth                     & North Dartmouth & USA\\\hline
P. Flin $\&$ J. Jalocha Bratek & Jan Kochanowski University   & Kielce     & Poland\\\hline
C. Flynn                & University of Sydney                & Sydney     & Australia\\\hline
C. Frigerio Martins     & Universidade Estadual de Campinas UNICAMP & Campinas & Brasil\\\hline
F. Frutos-Alfaro        & Space Research Center, University of Costa Rica  & San Jos\'{e} & Costa Rica\\\hline
J. Funes                & Specola Vaticana                    & Vatican    & Vatican\\\hline
P. Galianni             & University of St Andrews            & St Andrews & UK\\\hline
J. Gan                  & Shanghai Astronomical Observatory   & Shanghai   & China\\\hline
G. Gentile              & Katholieke Universiteit Leuven $\&$     & Leuven     & Belgium\\
                        & Vrije Universiteit Brussel          & Brussels   & Belgium\\\hline
I. George               & University Maryland Baltimore County& Baltimore County & USA\\\hline
S. George               & University of Calgary               & Calgary    & Canada\\\hline
L.\'{A}. Gergely $\&$ M. Dwornik & University of Szeged       & Szeged     & Hungary\\\hline 
M. Gomez                & Universidad de Huelva               & Huelva     & Spain\\\hline
P. Gondolo              & University of Utah                  & Salt Lake City & USA\\\hline
M. Goto                 & Universidade Estadual de Londrina   & Londrina   & Brazil\\\hline
M. Grossi               & University of Lisbon                & Lisbon     & Portugal\\\hline
L. T. Handoko           & Indonesian Institute of Sciences    & Tangerang  & Indonesia\\\hline 
T. Harko                & The University of Hong              & Hong Kong  & China\\\hline
N. Hashim, M.S.R. Hassan $\&$              & University of Malaya, Radio Cosmology Research Lab. & Kuala Lumpur & Malaysia\\
 I. Ungu &                                  &            & \\\hline
M. Hendry               & University of Glasgow               & Glasgow    & UK\\\hline
F. Hessman              & University of Goettingen            & G\"{o}ttingen & Germany\\\hline
S. Horvath              & University of Canterbury            & Christchurch& New Zealand\\\hline
D. Iakubovskyi          & Bogolyubov Institute for Theoretical Physics & Kiev & Ukraine\\\hline
C. Impey                & University of Arizona               & Tucson     & USA\\\hline
C. Kilinc               & Ege University                      & Izmir      & Turkey\\\hline
S. Kim                  & Kyung Hee University                & Seoul      & South Korea\\\hline
D. Kirilova  $\&$ G. Petrov & Bulgarian Academy of Sciences, Institute of Astronomy & Sofia & Bulgaria\\\hline
A. Knebe                & Universidad Autonoma de Madrid      & Madrid     & Spain\\\hline
S. Knollmann            & University of Zaragoza              & Zaragoza   & Spain\\\hline 
A. Lalovic $\&$ S. Samurovic & Astronomical Observatory in Belgrade & Belgrade & Serbia\\\hline
A. Lapi                 & University of Rome "Tor Vergata"    & Rome       & Italy\\\hline
B. Lazarotto Lago       & CEFET Nova Friburgo                 & Nova Friburgo & Brazil\\\hline
R. Machado              & IAG Universidade de S\~{A}o Paulo   & S\~{a}o Paulo & Brazil\\\hline
D. Makarov              & Russian Academy of Sciences         & Zelenchuck & Russia\\\hline
A. Maller               & The New York City College of Technology & New York & USA\\\hline 
G. Mamon                & Institut d'Astrophysique de Paris   & Paris      & France\\\hline 
D. Manreza Paret $\&$   & Universidad de La Habana            & La Habana  & Cuba\\
E. Rodriguez Querts     &                                     &            &\\\hline
C. Maraston             & Institute of Cosmology and Astrophysics & Portsmouth & UK\\\hline
D. Marchesini           & Tufts University                    & Medford    & USA\\\hline
C. Martins              & Centro de Astrofisica da Universidade do Porto   & Porto & Portugal\\\hline
T. Matos                & Cinvestav                           & Mexico City& Mexico\\\hline
T. Matos                & Instituto de F\'{i}sica y Matem\'{a}ticas de la Universidad Michoacana & Morelia & Mexico\\\hline
A. Meza                 & Universidad Andr\'{e}s Bello        & Santiago   & Chile\\\hline
S. Mohanty              & Physical Research Laboratory        & Ahmedabad  & India\\\hline
M. Molla                & Centro de Investigaciones Energ\'{e}ticas, Medioamb. y Tecnol. & Madrid & Spain\\\hline
J. C. Mu\~{n}oz Cuartas & Astrophysikalisches Institut Postdam& Potsdam    & Germany\\\hline
S. Murgia,  L. Strigari $\&$            & KIPAC, SLAC, Stanford University    & Stanford   & USA\\
 R. Wechsler &                                &            &\\\hline
K. Nagamine             & University of Nevada                & Las Vegas  & USA\\\hline
N. Napolitano           & Astronomical Observatory of Capodimonte & Naples & Italy\\\hline
D. Nieto Casta\~{n}o $\&$   & Complutense University of Madrid& Madrid     & Spain\\
J.A. Ruíz Cembranos     &                                     &            &\\\hline
F. Nesti                & INFN Laboratori Nazionali del Gran Sasso & L'Aquila & Italy\\\hline
A. Ogulenko             & Odessa I.I. Mechnikov National University & Odessa & Ukraine\\\hline
\'{A}. Orsi             & Universidad Cat\'{o}lica de Chile   & Santiago   & Chile\\\hline
N. Padilla              & Universidad Cat\'{o}lica de Chile   & Santiago   & Chile\\\hline
A. Parnowski            & Space Research Institute National Academy of Sciences & Kyiv & Ukraine\\\hline
A. Parnowski            & Kyiv Taras Shevchenko National University & Kyiv & Ukraine\\\hline
P. Patsis               & Academy of Athens, Research Center for Astronomy & Athens & Greece\\\hline
V. Pettorino            & Fourth TRR33 Winter School          & Passo del Tonale & Italy\\\hline
M. Plionis              & National Observatory of Athens, Inst. Astron. $\&$ Astrophys.& Athens & Greece\\\hline
J. Polednikova          & Faculty of Science, Masaryk University & Moravia & Czech Republic\\\hline
L. Reverberi            & University of Ferrara               & Ferrara    & Italy\\\hline
M. D. Rodriguez Frias   & University of Alcal\'{a}            & Alcal\'{a} de Henares & Spain\\\hline
A. Romeo                & Onsala Space Observatory, Chalmers & Onsala     & Sweden\\\hline
R. Rosenfeld            & Unesp Instituto de F\'{i}sica Te\'{o}rica & S\~{a}o Paulo & Brazil\\\hline
M. Roos                 & University of Helsinki $\&$              & Helsinki   & Finland\\
                        & University of Helsinki- Astrophysics Group & Helsinki & Finland\\\hline
A. Saburova $\&$ A. Zasov   & Sternberg Astronomical Institute    & Moscow     & Russia\\\hline
P. Salucci              & Observatory of Bologna              & Bologna    & Italy\\
                        & SISSA                               & Trieste    & Italy\\\hline
M. Sanchez-Conde        & Instituto de Astrofisica de Canarias& La Laguna  & Spain\\\hline
K. Sapountzis $\&$      & National $\&$ Kapodistrian University of Athens  & Athens & Greece\\
M. Petropoulou          &                                     &            &        \\\hline
M. Schneider            & Lawrence Livermore National Laboratory & Livermore & USA\\\hline
S. Schulze              & University of Iceland, Centre for Astrophysics and Cosmology & Dunhagi & Iceland\\\hline
D. Schwarz $\&$ M. Stuke    & Universitaet Bielefeld              & Bielefeld  & Germany\\\hline
M. Seigar               & University of Arkansas at Little Rock & Little Rock & USA\\\hline
N. Seymour              & Mullard Space Science Laboratory, University College London & Dorking &  UK\\\hline
M. D. Sheppeard         & Victoria University of Wellington   & Wellington & New Zealand\\\hline
Y-S.Song                & Korea Institute for Advanced Study  & Seoul      & Korea\\\hline
C. Struck               & Iowa State University               & Ames       & USA\\\hline
G. Tadaiesky Marques    & Universidade Federal do Par\'{a}    & Bel\'{e}m  & Brazil\\\hline
A. Tartaglia            & Politecnico di Torino, DIFIS        & Torino     & Italy\\\hline
A. N. Tawfik            & Egyption Center for Theoretical Physics & Cairo  & Egypt\\\hline
P. Thomas               & University of Sussex                & Sussex     & UK\\\hline
P. Tissera              & Institute for Astronomy and Space Physics & Buenos Aires & Argentina\\\hline
C. Tonini               & University of Melbourne             & Melbourne  & Australia\\\hline
L.A. Urena-Lopez        & University of Guanajuato, Physics Dept.& Guanajuato & Mexico\\\hline
J. van Eymeren          & Universitaet Duisburg-Essen $\&$        & Duisburg   & Germany\\
                        & Ruhr University Bochum              & Bochum     & Germany\\\hline
A. Velásquez-Toribio    & Universidade Federal de Juiz de Fora& Juiz de Forma & Brazil\\\hline
M. Viel                 & ICTP                                & Trieste    & Italy\\\hline
A. Vincent              & McGill University                   & Montreal   & Canada\\\hline
A-M. Weijmans $\&$ V. Sanz     & York University, Dept. Physics $\&$ Astronomy & York & Canada\\\hline
R. Wojtak               & Dark Cosmology Centre               & Copenhagen & Denmark\\\hline 
N. Yamasaki             & Institute of Space $\&$ Astronautical Science & Chofu & Japan\\\hline
F. Zandanel             & Instituto de Astrofisica de Andalucia& Granada   & Spain\\\hline      
B. Ziegler              & University of Vienna                & Vienna     & Austria\\ 
\hline
\end{longtable}
\end{center}

\end{document}